\documentclass[twocolumn,epjc3]{svjour3mod}

\usepackage{amsmath,amssymb,amsfonts,dcolumn,mathcomp,slashed,dsfont}
\usepackage{graphicx}
\usepackage{bm}
\usepackage{xcolor}
\usepackage{csquotes}
\usepackage{units}
\usepackage{flushend}
\usepackage{cite}
\usepackage[colorlinks,citecolor=blue,urlcolor=blue,linkcolor=blue]{hyperref}
\usepackage{ragged2e}
\usepackage{comment}
\usepackage{subcaption}
\usepackage{booktabs}
\usepackage{physics}

\apptocmd\thebibliography\justifying{}{}

\newcommand{\B}{\mathcal{B}}
\newcommand{\C}{\mathcal{C}}

\newcommand{\F}{\mathcal{F}}
\newcommand{\G}{\mathcal{G}}
\newcommand{\Ha}{\mathcal{H}}
\renewcommand{\L}{\mathcal{L}}
\newcommand{\M}{\mathcal{M}}
\newcommand{\N}{\mathcal{N}}
\renewcommand{\P}{\mathcal{P}}
\newcommand{\R}{\mathcal{R}}
\newcommand{\Q}{\mathcal{Q}}
\newcommand{\X}{\mathcal{X}}
\newcommand{\disc}{{\rm disc}\,}
\renewcommand{\dd}{{\rm d}}
\newcommand{\cf}{cf.}

\newcommand{\sth}{s_\text{th}}
\newcommand{\sIII}{s_\text{III}}
\newcommand{\sIV}{s_\text{IV}}

\newcommand{\GeV}{\,\text{GeV}}
\newcommand{\MeV}{\,\text{MeV}}
\newcommand{\bsp}{\begin{sloppypar}}
\newcommand{\esp}{\end{sloppypar}}

\graphicspath{{./plots/}{./figures/}{./plots/1--/}{./plots/1-+/}{./plots/0--/}}

\begin{document}

\title{\boldmath Analysis of rescattering effects in $3\pi$ final states}

\author{
Dominik Stamen\thanksref{Bonn,e1}
\and
Tobias Isken\thanksref{Bonn,GSI}
\and
Bastian Kubis\thanksref{Bonn,e2}
\and \\
Mikhail Mikhasenko\thanksref{Munich,e3}
\and
Malwin Niehus\thanksref{Bonn}
}

\thankstext{e1}{stamen@hiskp.uni-bonn.de}
\thankstext{e2}{kubis@hiskp.uni-bonn.de}
\thankstext{e3}{mikhail.mikhasenko@cern.ch}

\institute{Helmholtz-Institut f\"ur Strahlen- und Kernphysik (Theorie) and
   Bethe Center for Theoretical Physics,
   Universit\"at Bonn, 
   53115 Bonn, Germany \label{Bonn}
   \and
   Helmholtz Forschungsakademie Hessen für FAIR (HFHF) and GSI Helmholtzzentrum für Schwerionenforschung GmbH, Planckstraße 1, 64291 Darmstadt, Germany \label{GSI}
   \and
   ORIGINS Excellence Cluster and Ludwig-Maximilians-Universität München, 80939 Munich, Germany \label{Munich}
}

\date{}

\maketitle

\begin{abstract}
\bsp
Decays into three particles are often described in terms of two-body resonances and a non-interacting spectator particle.  To go beyond this simplest isobar model, crossed-channel rescattering effects need to be accounted for.
We quantify the importance of these rescattering effects in three-pion systems for different decay masses and angular-momentum quantum numbers.
We provide amplitude decompositions for four decay processes with total $J^{PC} = 0^{--}$, $1^{--}$, $1^{-+}$, and $2^{++}$, 
all of which decay predominantly as $\rho\pi$ states.
Two-pion rescattering is described in terms of an Omn{\`e}s function, which incorporates the $\rho$ resonance. 
Inclusion of crossed-channel effects is achieved by solving the Khuri--Treiman integral equations. 
The unbinned log-likelihood estimator is used to determine the significance of the rescattering effects beyond two-body resonances;
we compute the minimum number of events necessary to unambiguously find these in future Dalitz-plot analyses.
Kinematic effects that enhance or dilute the rescattering are identified for 
the selected set of quantum numbers and various masses.
\esp
\end{abstract}

\section{Introduction}\label{sec:introduction}

Much of the modern-era precision in hadron spectroscopy is not gleaned from scattering
reactions, but rather from production or decay processes~\cite{Amsler:1997up,Bugg:2004xu,Klempt:2007cp,Ketzer:2019wmd,Fang:2021wes}. 
For two-body states, the universality of final-state interactions~\cite{Watson:1954uc} provides an important
and fundamental link between scattering and production amplitudes, guaranteeing their phases to be
identical in the region where scattering is elastic. The presence of a third strongly interacting decay
product complicates rigorous analyses considerably.
An approximation to the decay amplitude where the two-particle interaction is modeled by a resonant amplitude without accounting for the interaction with the spectators is referred to as the isobar model.
The lineshape of the resonant pairwise interaction is often parameterized by the Breit--Wigner function~\cite{Breit:1936zzb}.
While one might expect that under certain circumstances, the influence of spectator particles on the
two-body resonance signal ought to be small---the resonance in question being narrow,
or the spectators having large relative momenta---the impact of the spectator interaction
on the resonance lineshape has only rarely been quantified. 
The goal of this article is to start the endeavor to survey such more complicated final-state interactions numerically,
beginning with the simplest processes: decays into three pions.

A tool to perform a theoretically rigorous evaluation of
three-body decays is given by the so-called Khuri--Treiman (KT) dispersion relations~\cite{Khuri:1960zz}.
These are coupled integral equations that describe all sequential pairwise two-body rescattering, summed to all orders.
While the two-body phase shifts are assumed to be known, the solutions of these equations depend on a set of free
parameters, subtraction constants of the dispersion integrals, which can be fixed by comparison to experimental
data or by matching to effective field theories. 
Instead of analyzing data for a particular process, we pose the question:
how much statistics needs to be collected in order for lineshape modifications due to a third pion to be discernible? 
We choose the processes to study based on the condition that their amplitude representations can be reduced to one single subtraction constant.
This then serves as a mere normalization, and hence allows us to study subtle variations of the resonance lineshape in an unambiguous manner.
In many three-pion decays, the $\rho$ meson is the most prominent pion--pion resonance,
and interactions of higher angular momentum are suppressed.
An important criterion is therefore that we select decays in which $S$-waves are forbidden
by conservation laws, reducing the problem to the $P$-wave interaction only.
One process of significant interest disregarded here is the $\tau\to 3\pi\nu_\tau$ decay, in which the $a_1(1260)$ resonance with $J^{PC}=1^{++}$ appears prominently; however, there are $S$-waves involved, whose strength relative to the $P$-waves cannot be fixed a priori without data (cf.\ a similar analysis in Ref.~\cite{Colangelo:2015kha}).

The outline of this article is as follows.  In Sec.~\ref{sec:KT}, we introduce and discuss the KT equations and derive the amplitude decompositions of four different three-pion decays that are dominated by the $\rho\pi$ intermediate states.  Subsequently, in Sec.~\ref{sec:loglikelihood}, we describe the statistical method used to distinguish the KT solutions from the simpler Omn{\`e}s model that neglects spectator interactions entirely.  Section~\ref{sec:results} shows our numerical results and discusses the dependence of the crossed-channel rescattering effects on mass and quantum numbers of the three-pion system in detail. 
Our findings are summarized in Sec.~\ref{sec:conclusions}.  Some technical details are relegated to the appendices.

\section{Khuri--Treiman equations}\label{sec:KT}

The Khuri--Treiman equations~\cite{Khuri:1960zz} were first derived in the 1960s to analyze $K\to 3\pi$ decays.
With the advent of very precise parameterizations of low-energy pion--pion ($\pi\pi$) phase shifts~\cite{Ananthanarayan:2000ht,Colangelo:2001df,Garcia-Martin:2011iqs}, the approach has experienced a remarkable renaissance and has been applied to various decays~\cite{Kambor:1995yc,Anisovich:1996tx,Niecknig:2012sj,Danilkin:2014cra,Guo:2015zqa,Niecknig:2015ija,Guo:2016wsi,Colangelo:2016jmc,Albaladejo:2017hhj,Isken:2017dkw,Niecknig:2017ylb,Colangelo:2018jxw,Dax:2018rvs,JPAC:2020umo,Akdag:2021efj,JPAC:2023nhq} and scattering processes~\cite{Hoferichter:2012pm,Hoferichter:2017ftn,Albaladejo:2018gif,Dax:2020dzg,Niehus:2021iin} since.

We here consider four different $3\pi$ decays with quantum numbers $J^{PC} = 0^{--}$, $1^{--}$, $1^{-+}$, and $2^{++}$, which fulfill the criterion introduced in Sec.~\ref{sec:introduction}.
For all these, the unpolarized distribution over the three-body phase space as represented by the Dalitz plot contains the full information on the decay dynamics, as only one helicity amplitude contributes in every case.
In order to apply the KT equations, we decompose each amplitude into the so-called single-variable amplitudes (SVAs), which are complex functions with a right-hand cut only. 
These decompositions are known as reconstruction theorems,
proven in chiral perturbation theory in a give order using fixed-variable dispersion relations~\cite{Stern:1993rg,Knecht:1995tr,Zdrahal:2008bd}.
Thereby, we restrict ourselves to $P$-waves and neglect all higher partial waves. 

\subsection[Reconstruction theorem for 1-- decay]{Reconstruction theorem for $1^{--}$ decay}

\begin{sloppypar}
We begin our discussion with the consideration of isoscalar vector quantum numbers, $I^G(J^{PC})=0^-(1^{--})$,
where the isospin $I$ is forced to be zero by the negative $G$-parity of the odd-pion system and negative $C$-parity.
Decays into three neutral pions are forbidden by charge conjugation.
The decays $\omega/\phi\to3\pi$ have been studied extensively using the KT formalism~\cite{Niecknig:2012sj,Danilkin:2014cra,Dax:2018rvs,JPAC:2020umo},
as well as extended to the $J/\psi\to3\pi$ decays~\cite{Kubis:2014gka}, and the general reactions $e^+e^-\to3\pi$~\cite{Hoferichter:2014vra,Hoferichter:2018dmo,Hoferichter:2018kwz,Hoferichter:2019mqg}. 
Experimentally, the Dalitz plots both for $\omega\to3\pi$~\cite{WASA-at-COSY:2016hfo,BESIII:2018yvu} and $\phi\to3\pi$~\cite{KLOE:2003kas,Akhmetshin:2006sc} have been investigated in detail.
\end{sloppypar}
The decay reads
\begin{align}
V(p)&\to \pi^0(p_1)\pi^+(p_2)\pi^-(p_3)\,,
\end{align}
where $p$ and $p_i$ denote the four momenta of the decay particle $V$ and the pions, respectively.
The Mandelstam variables~\cite{Mandelstam:1958xc} are defined as $s=(p-p_1)^2$, $t=(p-p_2)^2$, and $u=(p-p_3)^2$. The amplitude $\M$ is decomposed into a scalar amplitude $\F$ and a kinematic factor in the following form:
\begin{align}
\M(s,t,u)&=i\epsilon^\mu K_\mu \F(s,t,u)\,, \notag\\
K_\mu&=\varepsilon_{\mu\nu\alpha\beta}p_1^\nu p_2^\alpha p_3^\beta\,,\label{eq:MF_1--}
\end{align}
where the Levi-Civita tensor is employed due to the odd intrinsic parity.
By squaring the matrix element and averaging over the initial polarization, one obtains
\begin{align}
\overline{|\M|^2}&=\mathcal{K}(s,t,u)|\F(s,t,u)|^2\,,
\label{eq:M2_1--}
\end{align}
where $\mathcal{K}$ is a factor proportional to the Kibble function~\cite{Kibble:1960zz},
\begin{align} \label{eq:Kibble}
\mathcal{K}(s,t,u)&=\frac{1}{4}\left(stu-M_\pi^2(M^2-M_\pi^2)^2\right)\,.
\end{align}
Here, $M$ denotes the mass of the decay particle, and $M_\pi$ refers to the pion mass in the isospin limit.
Using the fixed-variable dispersion relations, one can show that the scalar amplitude is decomposed into the $P$-wave SVAs, $\F(x)$~\cite{Niecknig:2012sj,Niecknig:2016fva,Isken:2021gez,Niehus:2022aau}:
\begin{equation}
\F(s,t,u)=\F(s)+\F(t)+\F(u)\,, \label{eq:RT_1--}
\end{equation}
where discontinuities in $F$- and higher partial waves have been neglected.
The decay amplitude is invariant under a shift
\begin{equation}
\F(s) \to \F(s) + \alpha(s-s_0) \,,
\end{equation}
where $3s_0 \equiv M^2+3M_\pi^2 = s+t+u$, and $\alpha$ is an arbitrary complex constant.
This means that the decomposition~\eqref{eq:RT_1--} only defines $\F(s)$ up to such a polynomial ambiguity.
The strategy to eliminate the ambiguity is discussed in Sec.~\ref{sec:unitarity}.

\subsection[Reconstruction theorem for 1-+ decay]{Reconstruction theorem for $1^{-+}$ decay}\label{sec:rt_1mp}

Mesons with quantum numbers $I^G(J^{PC})=1^-(1^{-+})$ are exotic in the quark model.
The lightest candidate for such hybrid mesons is the $\pi_1(1600)$, which has been searched for experimentally in different final states such as $\eta\pi$~\cite{Schott:2012wqa,COMPASS:2014vkj,JPAC:2018zyd,CrystalBarrel:2019zqh,Kopf:2020yoa}, $\eta'\pi$~\cite{COMPASS:2014vkj,JPAC:2018zyd,Kopf:2020yoa}, and the most relevant for the present study $\rho\pi$~\cite{COMPASS:2018uzl}.

\begin{sloppypar}
The amplitude decomposition for the decays $\pi_1\to3\pi$ is very similar to the one for resonances with $I^G(J^{PC})=0^-(1^{--})$ discussed in the previous section, however, the positive charge conjugation implies odd isospin.
For the isovector decay, the explicit decomposition of the decay amplitude in terms of isospin indices is required,
\begin{align}
X^i(p)&\to \pi^j(p_1)\pi^k(p_2)\pi^l(p_3)\,,
\end{align}
where $i$, $j$, $k$, and $l$ are the isospin indices in Cartesian basis.
The decay is once more of odd intrinsic parity.
The decomposition reads
\begin{equation}
\M^{ijkl}(s,t,u)=i\epsilon^\mu K_\mu \mathcal{H}^{ijkl}(s,t,u)\,,\label{eq:amp_decomp}
\end{equation}
with the isospin amplitude $\mathcal{H}^{ijkl}$ following from the well-known isospin relations for $\pi\pi$ scattering~\cite{Weinberg:1966kf}:
\begin{align}
\mathcal{H}^{ijkl}(s,t,u)&=\delta^{ij}\delta^{kl}\mathcal{H}(s,t,u)+\delta^{ik}\delta^{jl}\mathcal{H}(t,u,s)&\notag\\
&+\delta^{il}\delta^{jk}\mathcal{H}(u,s,t)\,.
\label{eq:isospin_decomp}
\end{align}
The reconstruction theorem retaining $P$-waves only, i.e., neglecting discontinuities in $D$-waves and higher, reads
\begin{equation}
\mathcal{H}(s,t,u)=\mathcal{H}(t)-\mathcal{H}(u)\,,
\end{equation}
derived in \ref{app:1mp-RT}.
The decomposition is ambiguous by shifting
\begin{equation}
\mathcal{H}(s)\to\mathcal{H}(s)+\alpha\,. \label{eq:ambig_1mp}
\end{equation}
This ambiguity in principle allows us to write down a twice-subtracted dispersion integral for $\Ha(s)$ that still depends on one subtraction constant only. Since the subtraction constants of these two representations are connected by a sum rule, the amplitude $\Ha(s,t,u)$ remains unchanged.
The allowed charge configurations are
\begin{align}
X^+&\to\pi^+\pi^0\pi^0\notag\,,~~
X^+\to\pi^+\pi^+\pi^-\notag\,,~~ 
X^0\to\pi^+\pi^-\pi^0\notag\,,\\
X^-&\to\pi^-\pi^0\pi^0\notag\,,~~
X^-\to\pi^-\pi^-\pi^+\,,
\label{eq:charge_configs}
\end{align}
which all lead to the same result for the absolute squared of the amplitude.\footnote{The amplitudes for the charge configurations differ when including higher partial waves.} Due to the analogy between Eqs.~\eqref{eq:MF_1--} and \eqref{eq:amp_decomp}, the latter can be written in the same form as Eq.~\eqref{eq:M2_1--}. 
\end{sloppypar}

\subsection[Reconstruction theorem for 2++ decay]{Reconstruction theorem for $2^{++}$ decay}

For isovector tensor mesons, $I^G(J^{PC})=1^-(2^{++})$, the lightest state is the $a_2(1320)$ that dominantly decays into $3\pi$~\cite{ParticleDataGroup:2022pth}. Hence, we need to consider isospin explicitly, 
\begin{align}
T^i(p)&\to \pi^j(p_1)\pi^l(p_2)\pi^k(p_3)\,.
\end{align}
Due to the high spin of the decaying particle, the complete amplitude decomposition is considerably more complicated than for the vector decays and involves different helicity amplitudes; this is discussed in \ref{app:2pp-RT}.
The isospin decomposition of $\M^{ijkl}(s,t,u)$ involves invariant isospin amplitudes, which can be defined with respect to different Mandelstam variables. These are related to each other by crossing symmetry; see \ref{app:2pp-RT} for details. The $s$-channel amplitude is given as
\begin{align}
\M_s(s,t,u)&=i\sqrt{2}\epsilon_{\mu\nu}K^\mu \left[(p_2+p_3)^\nu \B(s,t,u) \right.\notag\\
&\left.\qquad+(p_2-p_3)^\nu \C(s,t,u)\right]\,.
\label{eq:amplitude_a2_mod}
\end{align}
The resulting spin-averaged squared amplitude is then given as
\begin{align}
\overline{|\M_s|^2}&=\tilde{\mathcal{K}}_1(s,t,u)|\B(s,t,u)|^2\notag\\
&\quad+2\tilde{\mathcal{K}}_2(s,t,u)\text{Re}\left(\B(s,t,u)\C(s,t,u)^*\right)\notag\\
&\quad+\tilde{\mathcal{K}}_3(s,t,u)|\C(s,t,u)|^2\,,
\label{eq:M2_2++}
\end{align}
where
\begin{align}
\tilde{\mathcal{K}}_i(s,t,u)&=\frac{\mathcal{K}(s,t,u)}{40M^2}k_i(s,t,u)\notag\,,\quad \forall i\in\{1,2,3\}\,,\\
k_1(s,t,u)&=\lambda(s,M^2,M_\pi^2)\notag\,,\\
k_2(s,t,u)&=(s+M^2-M_\pi^2)(u-t)\notag\,,\\
k_3(s,t,u)&=(t-u)^2+4M^2\left(s-4M_\pi^2\right)\,,
\end{align}
and $\lambda(a,b,c)=a^2+b^2+c^2-2(ab+ac+bc)$ is the standard K{\"a}ll{\'e}n function.
The reconstruction theorem of the scalar functions has the form
\begin{align}
\B(s,t,u)&=\B(t)-\B(u)\,,\notag\\
\C(s,t,u)&=\B(t)+\B(u)\,,
\end{align}
neglecting discontinuities that lead to $\pi\pi$ $D$-waves and higher.
This decomposition is unambiguous.

\subsection[Reconstruction theorem for 0-- decay]{Reconstruction theorem for $0^{--}$ decay}

The decays of the $\eta^{(\prime)}$ mesons, $I^G(J^{PC})=0^+(0^{-+})$, into $3\pi$ necessarily violate $G$-parity.
In the Standard Model, where both the strong and the electromagnetic interactions preserve charge conjugation, the decays proceed via breaking of isospin symmetry, while more exotic scenarios of physics beyond the Standard Model involving $C$-parity violation are suggested for the $\pi^+\pi^-\pi^0$ final state~\cite{Gardner:2019nid,Akdag:2021efj}.
We here concentrate on the latter, with total three-pion isospin $I=0$ and negative charge conjugation,
which would be equally applicable for the three-pion decay of a quark-model-exotic resonance with $I^G(J^{PC})=0^-(0^{--})$ in QCD.
Such states have been predicted as hybrid mesons with precisely this decay channel~\cite{Jaffe:1985qp}, although first lattice-QCD calculations at unphysically high pion masses suggest them to appear at higher masses than the $1^{-+}$ hybrids~\cite{Dudek:2011bn}; constituent-gluon models partly come to different conclusions~\cite{General:2006ed}.
The decay amplitude written in terms of $P$-waves only is given by
\begin{equation}
\M(s,t,u)=(t-u)\mathcal{G}(s)+(u-s)\mathcal{G}(t)+(s-t)\mathcal{G}(u)\,. \label{eq:RT_0--}
\end{equation}
We note that due to the decaying particle being a (pseudo)scalar, there is no additional kinematic factor in the relation to the Dalitz-plot distribution.
In contrast to the fully symmetric reconstruction theorem under pairwise exchange of Mandelstam variables for $J^{PC}=1^{--}$, this one is fully antisymmetric. 

The amplitude stays invariant under a three-parameter polynomial shift
\begin{equation}
\mathcal{G}(s)\to\mathcal{G}(s)+\alpha+\beta s+\gamma s^2(3s_0-s)\,,
\end{equation}
where $\alpha$, $\beta$, and $\gamma$ are arbitrary complex numbers.
Similar to the discussion in Sec.~\ref{sec:rt_1mp}, this ambiguity allows us to write $\mathcal{G}(s)$ as a twice- or three-times-subtracted dispersion integral, depending on a single subtraction constant.

\subsection{Partial-wave unitarity, Omn{\`e}s solutions}\label{sec:unitarity}

The form of the partial-wave series deviates due to the different spins of the decaying particles.
While the one for the pseudoscalar decay in Eq.~\eqref{eq:RT_0--} proceeds in terms of standard Legendre polynomials, the one for the vector decays 
has the form~\cite{Jacob:1959at}
\begin{equation}
\F(s,t,u)=\sum_{\ell=1}P_\ell'(z_s)f_\ell(s) \,,\label{eq:general_PWE_V}
\end{equation}
and similarly for $\Ha(s,t,u)$,
where $P_\ell'(z_s)$ refers to the derivatives of the Legendre polynomials.
The cosine of the $s$-channel scattering angle, denoted by $z_s$, can be expressed via the Mandelstam variables
\begin{align}
z_s=\frac{t-u}{\kappa(s)}\,, ~~ 
\kappa(s)=\sqrt{1-\frac{4M_\pi^2}{s}}\lambda^{1/2}(s,M^2,M_\pi^2)\,.\label{eq:scattangle}
\end{align}
In the $s$-channel center-of-mass system, $t$ and $u$ are related to the scattering angle via
\begin{align}
t(s,z_s)&=u(s,-z_s)=\frac{1}{2}\big(3s_0-s+\kappa(s)z_s\big) \,.\label{eq:t-u}
\end{align}
The partial-wave expansion for the tensor-meson decay is slightly more cumbersome as discussed in \ref{app:2pp-RT}.

\begin{figure}
    \centering
    \includegraphics[width=0.48\textwidth]{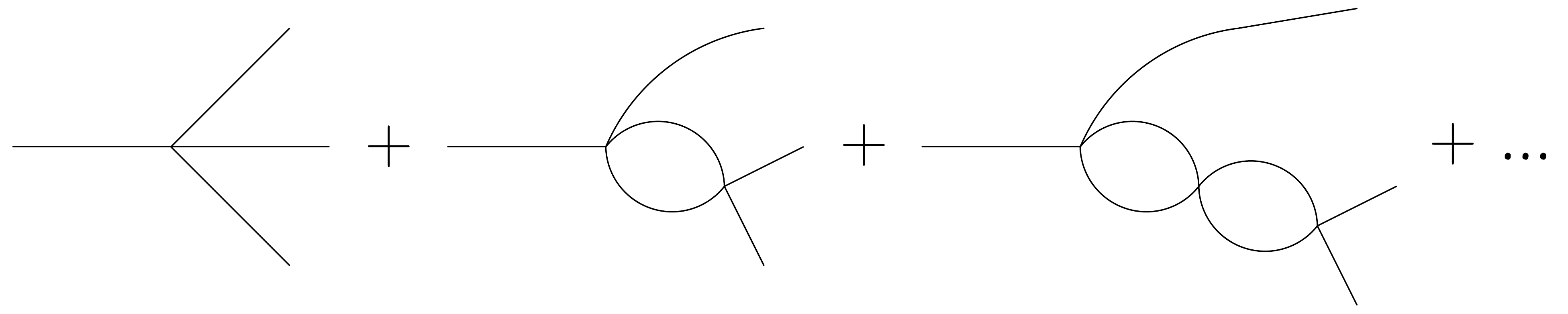}
    \caption{Diagrammatic representation of iterated bubble sums for the $2\pi$ subsystems, implemented by Omn{\`e}s functions. In general, due to interchange of the rescattered pions, three different bubble sums may contribute.}
    \label{fig:omnes_bubbles}
\end{figure}

\begin{figure}
    \centering
    \includegraphics[width=0.48\textwidth]{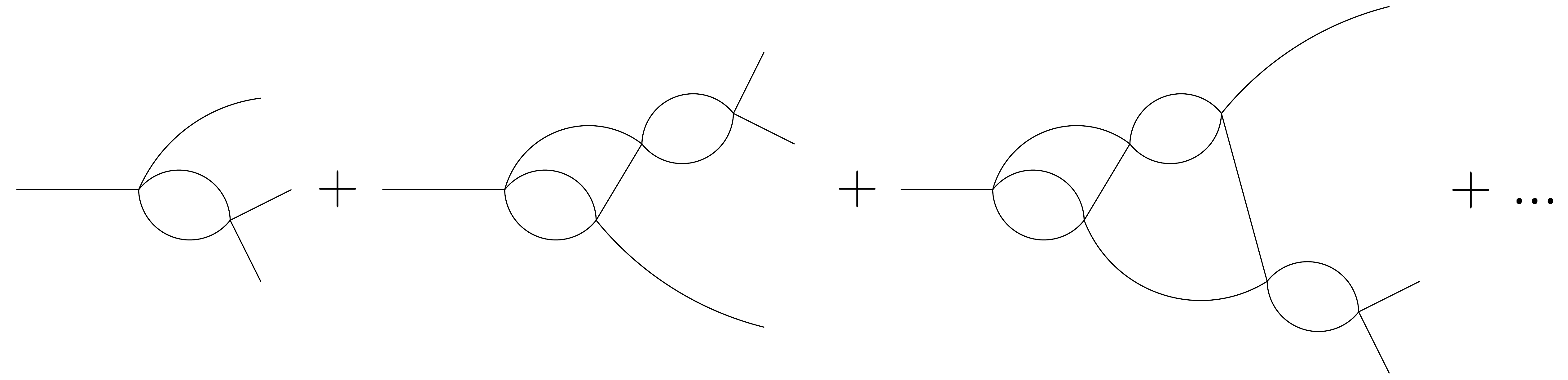}
    \caption{Diagrammatic representation of the amplitudes based on the full Khuri--Treiman equations.}
    \label{fig:crossed_bubbles}
\end{figure}

We consider elastic unitarity for the two-pion states. The $\pi\pi$ $P$-wave phase shift $\delta(s)=\delta_1^1(s)$ is parameterized according to Ref.~\cite{Colangelo:2018mtw}. 
A partial wave $\chi_1(s)$ of angular momentum $\ell=1$ obeys a unitarity relation of the form
\begin{align}
\disc \chi_1(s) &= 
\lim_{\epsilon\to0}\,[\chi_1(s+i\epsilon)-\chi_1(s-i\epsilon)]
\notag\\ &=
2i\chi_1(s)\sin\delta(s)e^{-i\delta(s)}\theta\big(s-4M_\pi^2\big)\,.
\end{align}
It can be decomposed into parts with right-hand and left-hand cuts only, $\chi_1(s)=\X(s)+\widehat{\X}(s)$, with $\X\in\{\F,\Ha,\B,\G\}$, where $\widehat{\X}$ is the so-called inhomogeneity that has no discontinuity along the right-hand cut.
It results from the partial-wave projection of the $t$- and $u$-channel SVAs.
The inhomogeneities are given by
\begin{alignat}{2}
\widehat{\F}(s)&=3\langle \left(1-z_s^2\right)\F\rangle && [J^{PC} = 1^{--}] \,, \notag\\
\widehat{\mathcal{H}}(s)&=-\frac{3}{2}\langle \left(1-z_s^2\right)\mathcal{H}\rangle && [J^{PC} = 1^{-+}] \,, \notag\\
\widehat{\B}(s)&=\frac{3}{4}\big[\langle\left(1-z_s^2\right)\B\rangle- \xi(s) &&\langle\left(1-z_s^2\right)z_s\B\rangle\big]\notag\\
& && [J^{PC} = 2^{++}] \,, \notag\\
\widehat{\G}(s)&=-\frac{3}{\kappa(s)}\Big[3(s-s_0)\langle z_s\G\rangle&&+\kappa(s)\langle z_s^2\G\rangle\Big] \notag\\
& && [J^{PC} = 0^{--}]\,,
\end{alignat}
where we employ the notation
\begin{equation}
\langle z_s^n \X\rangle = \frac{1}{2}\int_{-1}^1\dd z_s z_s^n \X\big(t(s,z_s)\big) \,,
\end{equation}
and $\xi(s)$ is defined via
\begin{equation}
\xi(s)=\sqrt{1-\frac{4M_\pi^2}{s}}\frac{s+M^2-M_\pi^2}{\lambda^{1/2}(s,M^2,M_\pi^2)}\,.
\end{equation}
As a consequence, the unitarity relations for the partial waves can be reduced to those for the SVAs $\X(s)$, which read
\begin{equation}
\disc \X(s)= 2i\Big(\X(s)+\widehat{\X}(s)\Big)\sin\delta(s)e^{-i\delta(s)}\theta\big(s-4M_\pi^2\big)\,.
\end{equation}
The solution for the homogeneous problem, setting $\widehat{\X}=0$, is given by the well-known Omn{\`e}s function $\Omega(s)$~\cite{Omnes:1958hv} 
\begin{align}
\X_{\text{hom.}}(s)&=P(s)\Omega(s)\,,\notag\\
\Omega(s)&=\exp\left(\frac{s}{\pi}\int_{4M_\pi^2}^\infty\dd s' \frac{\delta(s')}{s'(s'-s)}\right)\,,
\end{align}
where $P(s)$ is a polynomial and $\Omega(0)=1$. 
Such representations are used, e.g., in descriptions of the pion vector form factor; see Ref.~\cite{Colangelo:2018mtw} and references therein. 
Using the Omn{\`e}s function as an approximation for a SVA in a three-pion final state, we describe the rescattering of a two-pion subsystem only, with the third pion being a spectator, as shown in Fig.~\ref{fig:omnes_bubbles}. To include the full rescattering effects, cf.\ Fig.~\ref{fig:crossed_bubbles}, we need to solve the inhomogeneous equation. The solution is given by~\cite{Anisovich:1996tx}
\begin{align}
\X(s)&=\Omega(s)\Bigg(P_{n-1}(s)\notag\\
&\qquad+\frac{s^n}{\pi}\int_{4M_\pi^2}^\infty \frac{\dd s'}{s'^n}\frac{\sin\delta(s')\widehat{\X}(s')}{|\Omega(s')|(s'-s)}\Bigg)\,,
\label{eq:inhomOmnes-sol}
\end{align}
where $n$ determines the number of subtractions. 
As we aim for the most predictive model, without the need to fix the relative strength of various subtraction constants to concrete data, we set $n=1$. 
The high-energy behavior of the SVAs is dictated by the one of the Omn{\`e}s function, which in turn is given by the asymptotic limit of the input phase shift: $\delta(s\to\infty) \to \pi$ implies 
$\Omega(s\to\infty)\asymp s^{-1}$, and as a consequence, also the SVAs vanish asymptotically, $\X(s) \asymp s^{-1}$.
As a result, none of the polynomial ambiguities discussed in the previous sections survive: they would alter this asymptotic behavior and violate the high-energy constraint imposed.

We note that the mass of the decay particle enters $\widehat{\X}$ via the partial-wave projection integral.
Physically, $\widehat{\X}$ incorporates the crossed-channel effects, which depend on the relative momenta of all three final-state pions. The resulting differences are discussed in Sec.~\ref{sec:SVA}. The solution~\eqref{eq:inhomOmnes-sol} is generated iteratively for each process and decay mass, which can justify the diagrammatic representation in Fig.~\ref{fig:crossed_bubbles}. 

\begin{figure*}
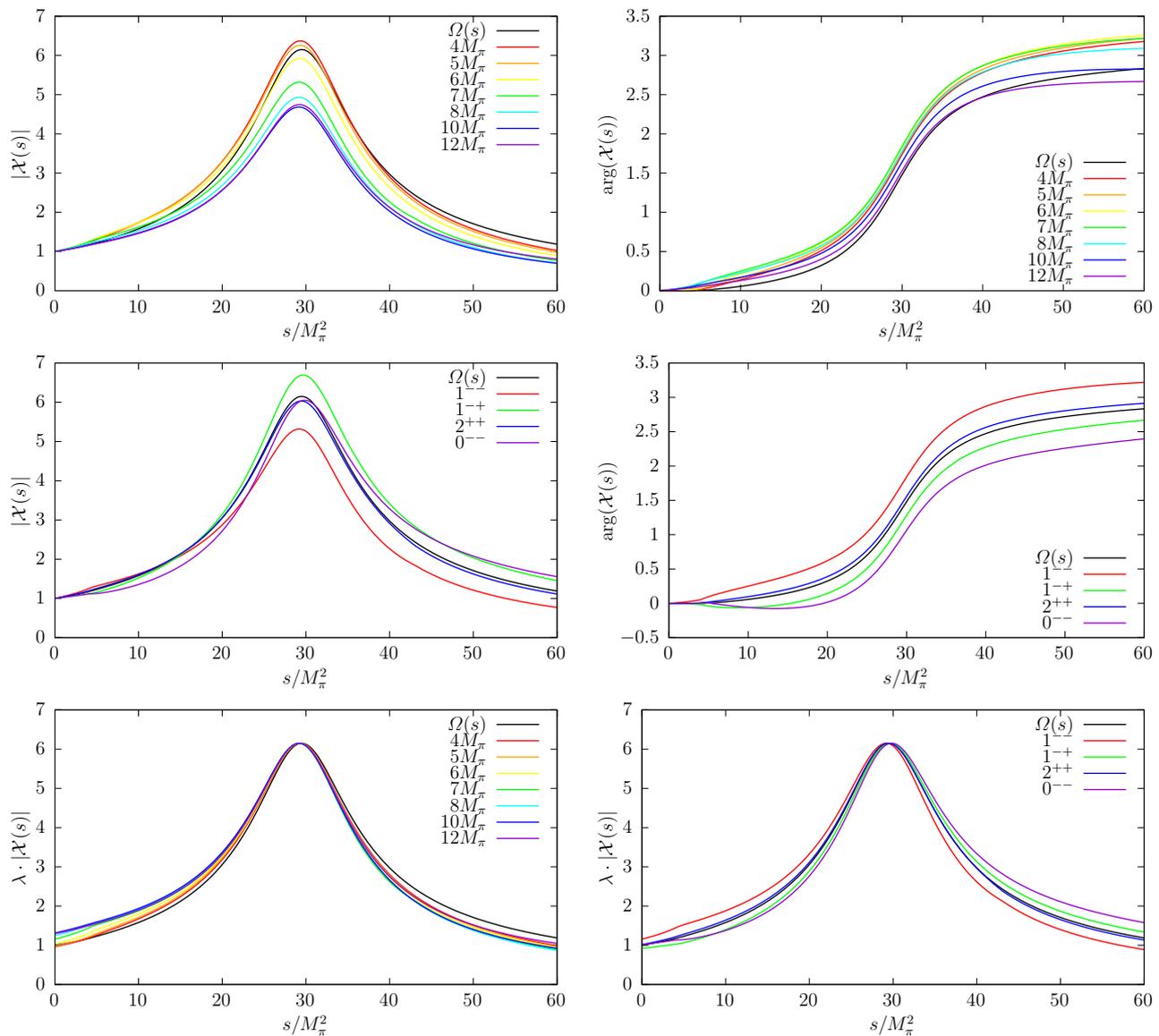

	\begin{subfigure}{0.49\textwidth}
	    \fontsize{12pt}{14pt} \selectfont
        \scalebox{0.67}{\input{plots/basis/basis_1mm_abs.tex}}
	\end{subfigure}
	\begin{subfigure}{0.49\textwidth}
    	\fontsize{12pt}{14pt} \selectfont
        \scalebox{0.67}{\input{plots/basis/basis_1mm_arg.tex}}
	\end{subfigure}
	
	\begin{subfigure}{0.49\textwidth}
	    \fontsize{12pt}{14pt} \selectfont
        \scalebox{0.67}{\input{plots/basis/basis_abs.tex}}
	\end{subfigure}
	\begin{subfigure}{0.49\textwidth}
    	\fontsize{12pt}{14pt} \selectfont
        \scalebox{0.67}{\input{plots/basis/basis_arg.tex}}
	\end{subfigure}
	
	\begin{subfigure}{0.49\textwidth}
	    \fontsize{12pt}{14pt} \selectfont
        \scalebox{0.67}{\input{plots/basis/basis_1mm_abs_norm.tex}}
	\end{subfigure}
	\begin{subfigure}{0.49\textwidth}
    	\fontsize{12pt}{14pt} \selectfont
        \scalebox{0.67}{\input{plots/basis/basis_abs_norm.tex}}
	\end{subfigure}

	\caption{Absolute values (left) and phases (right) of the SVAs: for $J^{PC}=1^{--}$ with different decay masses $M$ (top), and at the same decay mass $M=7M_\pi$ for the three different reconstruction theorems (middle). The two plots at the bottom show again the absolute values for both comparisons, but this time with the basis functions normalized to the peak of the Omn{\`e}s function.}
	\label{fig:basis}
\end{figure*}

\subsection{Comparison of different SVAs}\label{sec:SVA}

In Fig.~\ref{fig:basis}, we compare the KT solutions for the different SVAs to the Omn{\`e}s solution, both for different decay masses (for the $J^{PC}=1^{--}$ case; cf.\ also Ref.~\cite{Niecknig:2016fva}) and comparing the SVAs with different reconstruction theorems or inhomogeneities at the same decay mass. 
These pictures suggest we already have an answer to the question to what extent the $\rho$ lineshape and phase are modified by crossed-channel interactions, and how this modification varies with quantum numbers and decay mass.  However, this impression is misleading to some extent, as can be seen by the following considerations.
\begin{enumerate}
    \item As our reconstruction theorems all depend on one single SVA only, it is obvious that an overall shift of its phase by a constant is not observable.  A significant part of the changes in phase compared to the input phase shift seen in Fig.~\ref{fig:basis} can already be undone by such a shift.
    \item Although we have theoretically constrained our SVAs to fulfill a certain, restrictive, high-energy behavior, this still means that a polynomial shift according to the corresponding ambiguity is not observable in a finite Dalitz plot.  This suggests that any change between Omn{\`e}s and full KT solution that is, in fact, polynomial-like will not be experimentally verifiable.
    \item Finally, the single subtraction or normalization constants of our dispersive amplitude representations are not a priori fixed; changes in the SVAs that can be absorbed in a change of normalization will therefore also not allow us to verify non-trivial rescattering effects.  This is demonstrated in the bottom row of Fig.~\ref{fig:basis}, where the SVAs are not commonly normalized at $s=0$, but in the $\rho$ peak: the differences between the different solutions already appear significantly muted.
\end{enumerate}
All three points demonstrate that it is very difficult to quantify the observable changes by considering complex, interfering decay amplitudes only.  We therefore choose a different, unambiguous, path in the following and immediately study the Dalitz-plot distributions, which are direct observables.

\section{Log-likelihood estimator}\label{sec:loglikelihood}

In experiments measuring Dalitz plots, the data is often binned. The binning scheme is determined by each experiment individually to obtain distributions with reasonable statistical and systematic uncertainties; therefore, there is no unique prescription even for a given number of overall events. 
We thus seek an unbinned method to characterize the Dalitz-plot distribution, relying on the total number of events for the full Dalitz plot only. 
In the following, the deviations between the Omn{\`e}s and KT solutions are quantified using the Kullback--Leibler (KL) divergence~\cite{Kullback:1951zyt} based on the log-likelihood estimator.

As the description of the decay amplitudes in terms of KT equations is physically more complete, we interpret their 
outcome as the ``truth'' and test to which extent the Omn{\`e}s functions are capable of reproducing this. The decay amplitudes are denoted by $\M^\text{KT}$ and $\M^\text{Omn{\`e}s}$ for the KT and Omn{\`e}s solutions, respectively, where the isospin and helicity indices are suppressed for clarity. For the Omn{\`e}s solution, we replace the SVAs by plain Omn{\`e}s functions in the corresponding reconstruction theorems. 
We define the probability density function (pdf) as
\begin{align}
f(s,t)&=\frac{\overline{|\M|^2}(s,t,u)}{\int_D \overline{|\M|^2}(s,t,u)\dd s\dd t}\,.\label{eq:pdf}
\end{align}
Additionally, we explore an unphysical quantity defined by
\begin{align}
|\Tilde{\M}(s,t,u)|^2&=\frac{\overline{|\M|^2}(s,t,u)}{\mathcal{K}(s,t,u)} \,, \notag\\
\tilde{f}(s,t)&=\frac{|\Tilde{\M}(s,t,u)|^2}{\int_D |\Tilde{\M}(s,t,u)|^2\dd s\dd t}\,.\label{eq:pdf_nophase}
\end{align}
This construct appears to be more sensitive to the distribution at the edge of the Dalitz plots and therefore visualizes the $\rho$ bands discussed in Sec.~\ref{sec:results}.
Our representation is fixed up to a normalization constant, therefore the pdf is free of any undetermined parameters.
The integral region $D$ is the three-body decay phase space, captured by the Dalitz plot. For the decay mass $M$ and the pion mass $M_\pi$, its boundaries are determined by
\begin{align}
4M_\pi^2\leq s& \leq (M-M_\pi)^2\,,\notag\\
t(s,z_s=-1)\leq t& \leq t(s,z_s=1)\,,
\end{align}
where $t(s,z_s)$ is given by Eq.~\eqref{eq:t-u}.
We generate a sample $\mathbb{D}$ by drawing $N\in\mathbb{N}$ pairs $(s_i,t_i)$ from $f^\text{KT}$. The likelihood function $L$ of a pdf $f$ with respect to the data sample $\mathbb{D}$ is defined as
\begin{equation}
L(\mathbb{D},f)=\prod_{i=1}^N f(s_i,t_i)\,.
\end{equation}
The likelihood ratio of the Omn{\`e}s and KT solutions
\begin{equation}
\Delta L(\mathbb{D})=\frac{L(\mathbb{D},f^\text{Omn{\`e}s})}{L(\mathbb{D},f^\text{KT})} 
\end{equation}
indicates which one is favored.
In the following, we will use the log-likelihood and its difference
\begin{align}
\L(\mathbb{D},f)&=\ln(L(\mathbb{D},f))\,,\notag\\
\Delta \L(\mathbb{D}) &= \ln\left(\Delta L (\mathbb{D})\right)=\L(\mathbb{D}, f^\text{Omn{\`e}s})-\L(\mathbb{D},f^\text{KT})\,.
\end{align}
We note that $\Delta\L>0$ is possible despite drawing data from the KT solution, since $N$ is finite. This gives us precisely the handle we need to determine the value of $N$ to observe crossed-channel rescattering effects.

\begin{figure}
    \fontsize{12pt}{14pt} \selectfont
    \scalebox{0.66}{\input{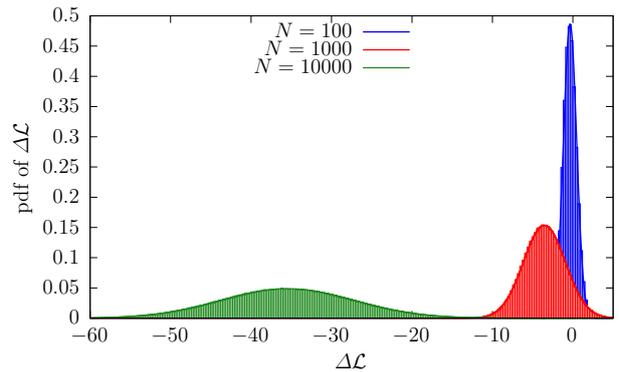}}
    \caption{Histogram for $B=10^6$ datasets of different sample size $N$. For this plot, the amplitudes are computed at $M=M_\phi$ and for $J^{PC}=1^{--}$. Additionally, we plot Gaussians with the mean and standard deviation from Eq.~\eqref{eq:mean_std}.}
    \label{fig:histo}
\end{figure}

We can now perform $B\in\mathbb{N}$ runs, which generate $B$ datasets $\mathbb{D}_b$, $b=1,\ldots,B$, of size $N$. On each of these datasets, one can compute $\Delta\L$ and access its probabilistic distribution.
For large values of $B$, this distribution is Gaussian, with the mean and variance given by
\begin{align}
\text{E}\left[\Delta\L(\mathbb{D})\right]&=-N d_\text{KL} \,,\notag\\
\text{Var}\left[\Delta\L(\mathbb{D})\right]&=N\nu_\text{KL}\,,
\end{align}
where
\begin{align}
\tilde{d}_\text{KL}(s,t)&=f^\text{KT}(s,t) \ln\left(\frac{f^\text{KT}(s,t)}{f^\text{Omn{\`e}s}(s,t)}\right)\,,\notag\\
d_\text{KL}&=\int_D \tilde{d}_\text{KL}(s,t)\dd s\dd t \,,\notag\\
\tilde{\nu}_\text{KL}(s,t)&=f^\text{KT}(s,t) \ln\left(\frac{f^\text{KT}(s,t)}{f^\text{Omn{\`e}s}(s,t)}\right)^2\,,\notag\\
\nu_\text{KL}&= \int_D \tilde{\nu}_\text{KL}(s,t) \dd s\dd t - d^2_\text{KL}\,.
\label{eq:entropy_variance}
\end{align}
The expressions are known as the Kullback--Leibler divergence~\cite{Kullback:1951zyt} and variance.
The cumulative distribution function reads
\begin{align}
\N(x,\mu(N),\sigma(N))&=\frac{1}{2}\left(1+\text{erf}\left(\frac{x-\mu(N)}{\sqrt{2}\sigma(N)}\right)\right) \notag\\
\text{with} \quad \mu(N)&=-Nd_\text{KL}\,, \quad \sigma(N)=\sqrt{N\nu_\text{KL}}\,,
\label{eq:mean_std}
\end{align}
where erf is the error function. To validate the assumption of a normal distribution, we use the comparison in Fig.~\ref{fig:histo}, which indicates a very good description.\footnote{Note that this is not an assumption for large $N$ due to the central-limit theorem.}
From here on we can calculate our results using the pdfs as defined in Eq.~\eqref{eq:pdf}.

In the region $\Delta\L<0$ we reject the hypothesis that the Omn{\`e}s solutions are sufficient to describe the data. 
The probability that the hypothesis is not rejected then reads
\begin{equation}
q(N)=1-\N(0,\mu(N),\sigma(N))\,.
\end{equation}
The inversion of the equation gives the number of events with the confidence determined by $q$ via
\begin{equation}
N(q)=2\nu_\text{KL}\left(\frac{\text{erf}^{-1}(1-2q)}{d_\text{KL}}\right)^2\,.
\end{equation}
For a $5\sigma$ confidence level we need to have $q=2.87\cdot10^{-7}$~\cite{ParticleDataGroup:2022pth} and can now compute the resulting $N$.

\begin{figure*}[t!]
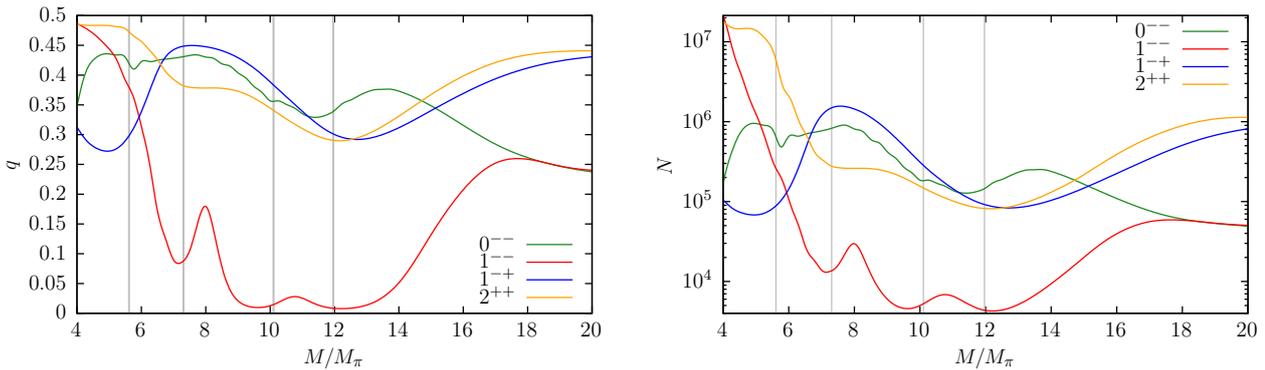

    \begin{subfigure}{0.49\textwidth}
	    \fontsize{12.5pt}{14pt} \selectfont
        \scalebox{0.65}{\input{plots/q_mass_dependence.tex}}
	\end{subfigure}
    \begin{subfigure}{0.49\textwidth}
	    \fontsize{12.5pt}{14pt} \selectfont
        \scalebox{0.65}{\input{plots/N_mass_dependence.tex}}
	\end{subfigure}
    \caption{\textit{Left panel:} Probability of the Omn{\`e}s model not being rejected as a function of the mass of the decaying particle, for all four reconstruction theorems and $N=1000$. \textit{Right panel:} Statistics needed to set the probability that the Omn{\`e}s model is not rejected to $5\sigma$ (right). The gray vertical lines, from left to right correspond to the masses of the $\omega$, $\phi$, $\omega(1420)$, and $\omega(1650)$ resonances.}
    \label{fig:mass_dep}
\end{figure*}

\begin{figure*}[t!]
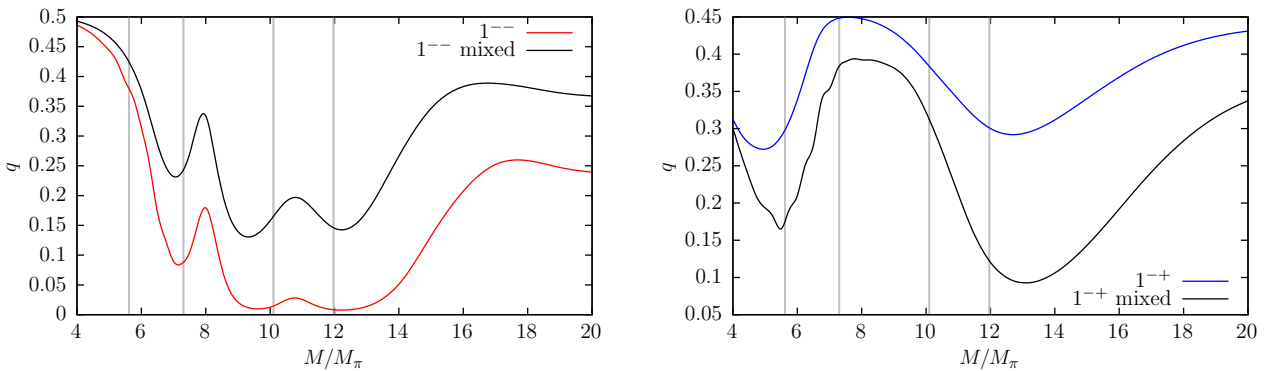

    \begin{subfigure}{0.49\textwidth}
	    \fontsize{12.5pt}{14pt} \selectfont
        \scalebox{0.65}{\input{plots/q_mass_dependence_diff_1mm.tex}}
	\end{subfigure}
    \begin{subfigure}{0.49\textwidth}
	    \fontsize{12.5pt}{14pt} \selectfont
        \scalebox{0.65}{\input{plots/q_mass_dependence_diff_1mp.tex}}
	\end{subfigure}
    \caption{\textit{Left panel:} Comparison for the $1^{--}$ reconstruction theorem with normal SVAs and $1^{-+}$ SVAs (``mixed''). \textit{Right panel:} Comparison for $1^{-+}$ reconstruction theorem with normal SVAs and $1^{--}$ SVAs (``mixed''). We show the probability of the Omn{\`e}s model not being rejected as a function of the mass of the decaying particle. Vertical gray lines as in Fig.~\ref{fig:mass_dep}.}
    \label{fig:q_mass_diff}
\end{figure*}

\section{Results}\label{sec:results}

Among the three-pion decays studied in this paper, only two Dalitz plots have been studied experimentally with sufficiently high statistics: $\omega\to3\pi$~\cite{WASA-at-COSY:2016hfo,BESIII:2018yvu}, and $\phi\to 3\pi$~\cite{KLOE:2003kas,Akhmetshin:2006sc}.

For $\omega\to 3\pi$, WASA-at-COSY~\cite{WASA-at-COSY:2016hfo} has performed a Dalitz plot study with $44\,080$ events, 
while the analysis by BESIII~\cite{BESIII:2018yvu} is based on $260\,520$ events. 
Both experiments parameterize the distribution by a polynomial expansion and present results testing one- and two-parameter models.
Applying the formalism of the preceding section, we find that the statistics of WASA-at-COSY is sensitive to rescattering effects only at $2.1\sigma$.  On the other hand, in agreement with Ref.~\cite{JPAC:2020umo}, BESIII reaches a $5\sigma$ level for the solution containing one subtraction.
However, as pointed out by Ref.~\cite{JPAC:2020umo}, an additional subtraction leads to a better agreement for the Dalitz plot parameters. 

For $\phi\to 3\pi$,  KLOE~\cite{KLOE:2003kas} provides a Dalitz plot analysis using $2\cdot 10^6$ events, while CMD-2~\cite{Akhmetshin:2006sc} has studied almost $8\cdot 10^4$ decays. For both, rescattering effects are clearly observable, as concluded by Ref.~\cite{Niecknig:2012sj}. 

Using the statistical method explained above and the derived reconstruction theorems, we compute $q(N)$ and $N(q)$ for a large mass range as shown in Fig.~\ref{fig:mass_dep}. The mass dependence of $N$ and $q$ for the $1^{--}$ decay looks strikingly different from the $1^{-+}$ one, even though they share the same kinematic factor.
To investigate the source of this difference in sensitivity between the two reconstruction theorems, we perform the following, unphysical, test.  We plug the SVAs, calculated as KT solutions for the $1^{-+}$ decay, into the linear combination given by the reconstruction theorem for $1^{--}$, see Eq.~\eqref{eq:RT_1--}, and vice versa.
These unphysical amplitudes are denoted by ``mixed'' in Fig.~\ref{fig:q_mass_diff}. We observe that this changes the absolute values of $q(N)$, while the qualitative behavior is the same.  We therefore conclude that much of the sensitivity to rescattering effects is not actually due to the size of the lineshape modification of the SVAs, as shown in Fig.~\ref{fig:basis}, but rather due to the specific linear combination in which they form the full decay amplitude.

In a log-plot for $N$ (for fixed $q$) as a function of the decay mass, we find a similar form as for $q$ (with fixed $N$); cf.\ left and right panels of Fig.~\ref{fig:mass_dep}. For large decay masses, the necessary number of events rises for all processes.\footnote{Note that this effect is not fully visualized by the mass range displayed in Fig.~\ref{fig:mass_dep}.} This is due to the fact that the KT solutions converge to the Omn{\`e}s function in the infinite-mass limit.
However, in the high-mass region, $M\gtrsim 15M_\pi$, our approximations are no longer valid: inelastic effects and higher partial waves play a non-negligible role. For low decay masses, approaching the three-pion threshold, the necessary event numbers for the $1^{--}$, $1^{-+}$, and $2^{++}$ decays rise due to limited phase space and kinematic suppression of the Dalitz-plot borders, far away from the $\rho$ resonance. For $0^{--}$ this is different, since here the amplitude does not vanish at the edge of the Dalitz plot.

\begin{figure*}
	\begin{subfigure}{0.49\textwidth}
	    \fontsize{12.5pt}{14pt} \selectfont
        \scalebox{0.67}{\input{plots/1-+/dal_5.0.tex}}
	\end{subfigure}
	\begin{subfigure}{0.49\textwidth}
    	\fontsize{12.5pt}{14pt} \selectfont
        \scalebox{0.67}{\input{plots/1-+/dal_comp_5.0.tex}}
	\end{subfigure}
	
	\begin{subfigure}{0.49\textwidth}
    	\fontsize{12.5pt}{14pt} \selectfont
        \scalebox{0.67}{\input{plots/1-+/dal_7.25.tex}}
	\end{subfigure}
	\begin{subfigure}{0.49\textwidth}
        \fontsize{12.5pt}{14pt} \selectfont
        \scalebox{0.67}{\input{plots/1-+/dal_comp_7.25.tex}}
	\end{subfigure}

	\begin{subfigure}{0.49\textwidth}
		\fontsize{12.5pt}{14pt} \selectfont
        \scalebox{0.67}{\input{plots/1-+/dal_8.0.tex}}
	\end{subfigure}
	\begin{subfigure}{0.49\textwidth}
		\fontsize{12.5pt}{14pt} \selectfont
        \scalebox{0.67}{\input{plots/1-+/dal_comp_8.0.tex}}
	\end{subfigure}
	
	\begin{subfigure}{0.49\textwidth}
		\fontsize{12.5pt}{14pt} \selectfont
        \scalebox{0.67}{\input{plots/1-+/dal_14.0.tex}}
	\end{subfigure}
	\begin{subfigure}{0.49\textwidth}
		\fontsize{12.5pt}{14pt} \selectfont
        \scalebox{0.67}{\input{plots/1-+/dal_comp_14.0.tex}}
	\end{subfigure}
	\caption{Dalitz plots for the $J^{PC}=1^{-+}$ reconstruction theorem. From top to bottom the decay mass grows according to $M=5/7.25/8/14\,M_\pi$, while the Dalitz plot without phase space $\tilde{f}(s,t)$ defined via Eq.~\eqref{eq:pdf_nophase} is shown in the left and $\tilde{d}_\text{KL}(s,t)$ (Eq.~\eqref{eq:entropy_variance}) in the right column.}
	\label{fig:dalitz_1mp}
\end{figure*}

The mass scan manifests several prominent features in the significance plot. 
For $1^{-+}$ decays, $N$ starts at $10^5$ events around the $\omega$ mass and then rises steeply to approximately $2\cdot10^6$ events at the $\phi$ mass.  At higher masses, it falls off up to about $12M_\pi$. The very high number of necessary events is mainly due to the fact that crossing symmetry requires a zero in the Dalitz plot along the line $t=u$, and hence any differences due to rescattering have to appear at the edge of the Dalitz plot, where the phase space is suppressed by the Kibble function. The Dalitz plots for decays of a particle with $1^{-+}$ quantum numbers are shown in Fig.~\ref{fig:dalitz_1mp} for different masses. Here the difference decreases 
until the $\rho$ bands are inside the Dalitz plot, and then falls off again when the size increases further.

\begin{figure*}
	\begin{subfigure}{0.49\textwidth}
	    \fontsize{12.5pt}{14pt} \selectfont
        \scalebox{0.67}{\input{plots/1--/dal_5.0.tex}}
	\end{subfigure}
	\begin{subfigure}{0.49\textwidth}
    	\fontsize{12.5pt}{14pt} \selectfont
        \scalebox{0.67}{\input{plots/1--/dal_comp_5.0.tex}}
	\end{subfigure}
	
	\begin{subfigure}{0.49\textwidth}
    	\fontsize{12.5pt}{14pt} \selectfont
        \scalebox{0.67}{\input{plots/1--/dal_7.25.tex}}
	\end{subfigure}
	\begin{subfigure}{0.49\textwidth}
        \fontsize{12.5pt}{14pt} \selectfont
        \scalebox{0.67}{\input{plots/1--/dal_comp_7.25.tex}}
	\end{subfigure}

	\begin{subfigure}{0.49\textwidth}
		\fontsize{12.5pt}{14pt} \selectfont
        \scalebox{0.67}{\input{plots/1--/dal_8.0.tex}}
	\end{subfigure}
	\begin{subfigure}{0.49\textwidth}
		\fontsize{12.5pt}{14pt} \selectfont
        \scalebox{0.67}{\input{plots/1--/dal_comp_8.0.tex}}
	\end{subfigure}
	
	\begin{subfigure}{0.49\textwidth}
		\fontsize{12.5pt}{14pt} \selectfont
        \scalebox{0.67}{\input{plots/1--/dal_14.0.tex}}
	\end{subfigure}
	\begin{subfigure}{0.49\textwidth}
		\fontsize{12.5pt}{14pt} \selectfont
        \scalebox{0.67}{\input{plots/1--/dal_comp_14.0.tex}}
	\end{subfigure}
	\caption{Dalitz plots for the $J^{PC}=1^{--}$ reconstruction theorem. From top to bottom the decay mass grows according to $M=5/7.25/8/14\,M_\pi$, while the Dalitz plot without phase space $\tilde{f}(s,t)$ defined via Eq.~\eqref{eq:pdf_nophase} is shown in the left and $\tilde{d}_\text{KL}(s,t)$ (Eq.~\eqref{eq:entropy_variance}) in the right column.} 
	\label{fig:dalitz_1mm}
\end{figure*}

\begin{figure}
    \centering
    \includegraphics[width=0.48\textwidth]{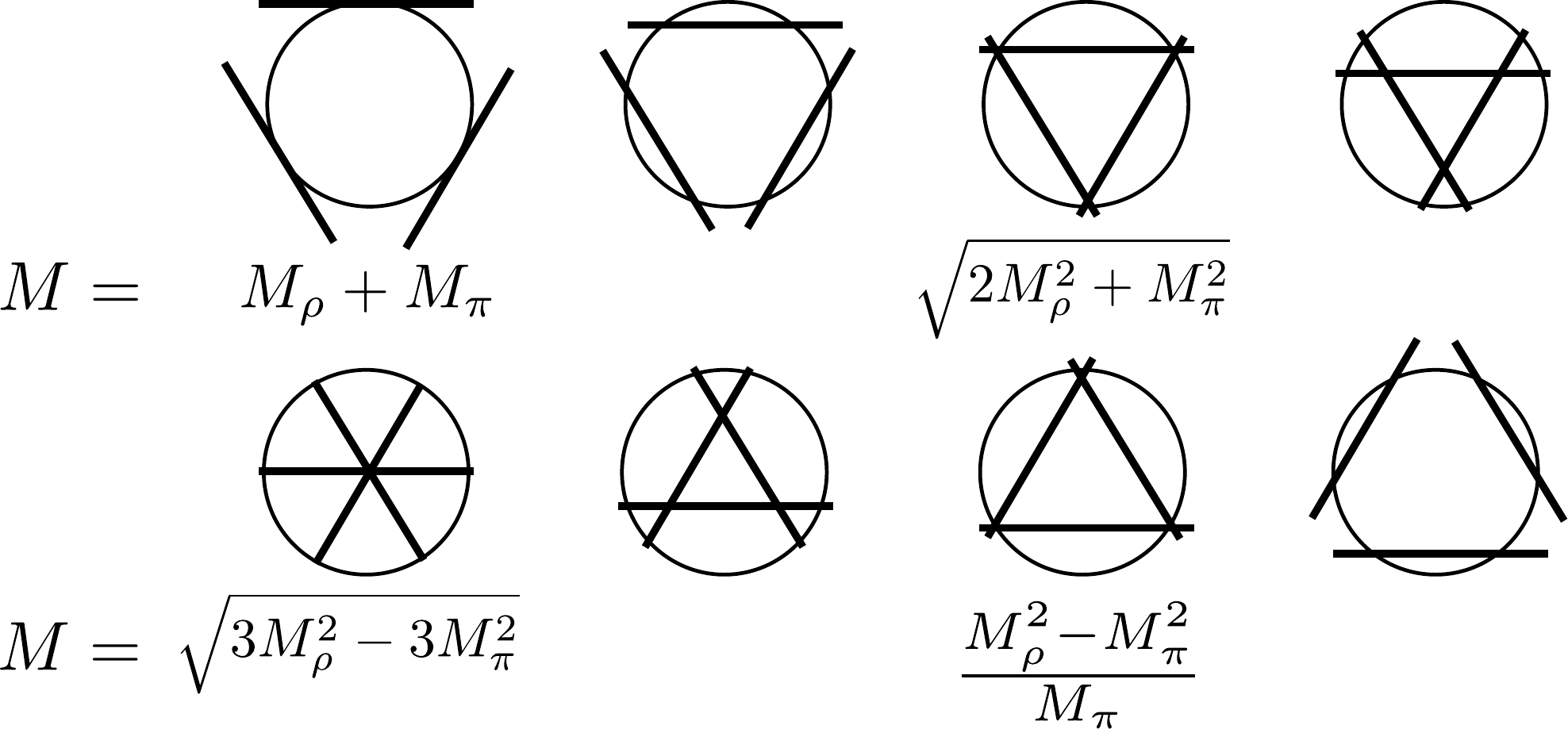}
    \caption{Sketches of different kinematic configurations for the $1^{--}$ decay with increasing decay mass.  The straight lines denote the (qualitative) position of the three $\rho$ bands in the Dalitz plot.  The masses below the diagrams denote the decay masses for which the specific kinematic configurations are reached.}
    \label{fig:configurations}
\end{figure}

For the $1^{--}$ decays, we find a different behavior. The event number $N$ starts at high values for the $\omega$ resonance and then shows an overall decline with rising decay mass, with two small peaks between the $\phi$ and $\omega(1650)$. At the $\phi$ resonance mass, the $\rho$ bands are completely inside the Dalitz plot. The Dalitz plots for different decay masses are shown in Fig.~\ref{fig:dalitz_1mm}. The first peak occurs due to the third kinematic configuration of the Dalitz plot as shown in Fig.~\ref{fig:configurations}. The difference increases again when the three $\rho$ bands cross in the middle of the Dalitz plot. The second peak is also due to a peculiar structure in the Dalitz plot: in this decay mass region we find a ring-shaped local minimum, clearly visible for an unphysically narrow $\rho$ width; see \ref{app:interferencering}.  
The ring affects the sensitivity even for the physical $\rho$ width, and is responsible for the second peak.

\begin{figure}
    \centering
    \fontsize{12pt}{14pt} \selectfont
    \scalebox{0.65}{\input{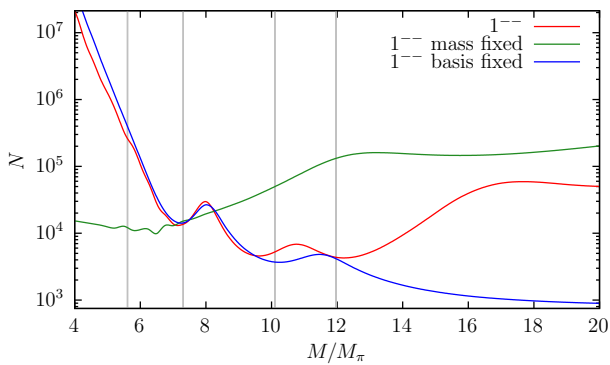}}
    \caption{Minimal number of events $N$ to exclude the Omn{\`e}s model at $5\sigma$ significance for $J^{PC}=1^{--}$, as a function of the decay mass decomposed into two different effects.
    The red line corresponds to the full solution from the right panel in Fig.~\ref{fig:mass_dep}. For the green line (``mass fixed''), we fix the size of the Dalitz plot to $M=7.5M_\pi$ and only use the SVA varying with the running mass. For the blue line (``basis fixed''), we use the same SVA solution for $M=7.5M_\pi$ for all masses and vary the size of the Dalitz plot.
    Vertical gray lines as in Fig.~\ref{fig:mass_dep}.}
    \label{fig:N_mass_fixed_1mm}
\end{figure}

In order to disentangle the origin of the various maxima and minima in the sensitivity of the $1^{--}$ decays in dependence on the decay mass a little better, we separate, somewhat unphysically, two different effects in Fig.~\ref{fig:N_mass_fixed_1mm}: the size of the Dalitz plot, and modifications of the SVAs.  We once keep the SVA basis function fixed as calculated for decay mass $M=7.5M_\pi$ and only vary the size of the Dalitz plot; and secondly, we vice versa keep the Dalitz plot fixed at $M=7.5M_\pi$, and only vary the SVA with its implicit decay-mass dependence.  
The precise plot depends heavily on the choice for the fixed mass, but we observe both peaks when using a SVA for a fixed mass and only varying the size of the Dalitz plot.  
Fixing the size of the Dalitz plot and only varying the SVAs, on the other hand, results in a rather smooth decrease of the difference towards higher masses. We therefore conclude that the peaks in the $1^{--}$ mass dependence are dominated by the structure of the Dalitz plot and not by the difference in the SVAs.

The main takeaway however is that with less than $10^5$ events above the $\phi$ mass, rescattering effects will play an important role in analyses of Dalitz plot data. 
For the $\omega$ resonance and lower masses of the decaying particle, one requires more than $10^6$ events to observe them.

\begin{figure*}
	\begin{subfigure}{0.49\textwidth}
	    \fontsize{12.5pt}{14pt} \selectfont
        \scalebox{0.67}{\input{plots/0--/dal_5.0.tex}}
	\end{subfigure}
	\begin{subfigure}{0.49\textwidth}
    	\fontsize{12.5pt}{14pt} \selectfont
        \scalebox{0.67}{\input{plots/0--/dal_comp_5.0.tex}}
	\end{subfigure}
	
	\begin{subfigure}{0.49\textwidth}
    	\fontsize{12.5pt}{14pt} \selectfont
        \scalebox{0.67}{\input{plots/0--/dal_7.25.tex}}
	\end{subfigure}
	\begin{subfigure}{0.49\textwidth}
        \fontsize{12.5pt}{14pt} \selectfont
        \scalebox{0.67}{\input{plots/0--/dal_comp_7.25.tex}}
	\end{subfigure}

	\begin{subfigure}{0.49\textwidth}
		\fontsize{12.5pt}{14pt} \selectfont
        \scalebox{0.67}{\input{plots/0--/dal_8.0.tex}}
	\end{subfigure}
	\begin{subfigure}{0.49\textwidth}
		\fontsize{12.5pt}{14pt} \selectfont
        \scalebox{0.67}{\input{plots/0--/dal_comp_8.0.tex}}
	\end{subfigure}
	
	\begin{subfigure}{0.49\textwidth}
		\fontsize{12.5pt}{14pt} \selectfont
        \scalebox{0.67}{\input{plots/0--/dal_14.0.tex}}
	\end{subfigure}
	\begin{subfigure}{0.49\textwidth}
		\fontsize{14pt}{14pt} \selectfont
        \scalebox{0.67}{\input{plots/0--/dal_comp_14.0.tex}}
	\end{subfigure}
	\caption{Dalitz plots for the $J^{PC}=0^{--}$ reconstruction theorem. From top to bottom the decay mass grows according to $M=5/7.25/8/14\,M_\pi$, while the Dalitz plot $f(s,t)$ defined via Eq.~\eqref{eq:pdf} is shown in the left and $\tilde{d}_\text{KL}(s,t)$ (Eq.~\eqref{eq:entropy_variance}) in the right column.} 
	\label{fig:dalitz_0mm}
\end{figure*}

\begin{figure*}
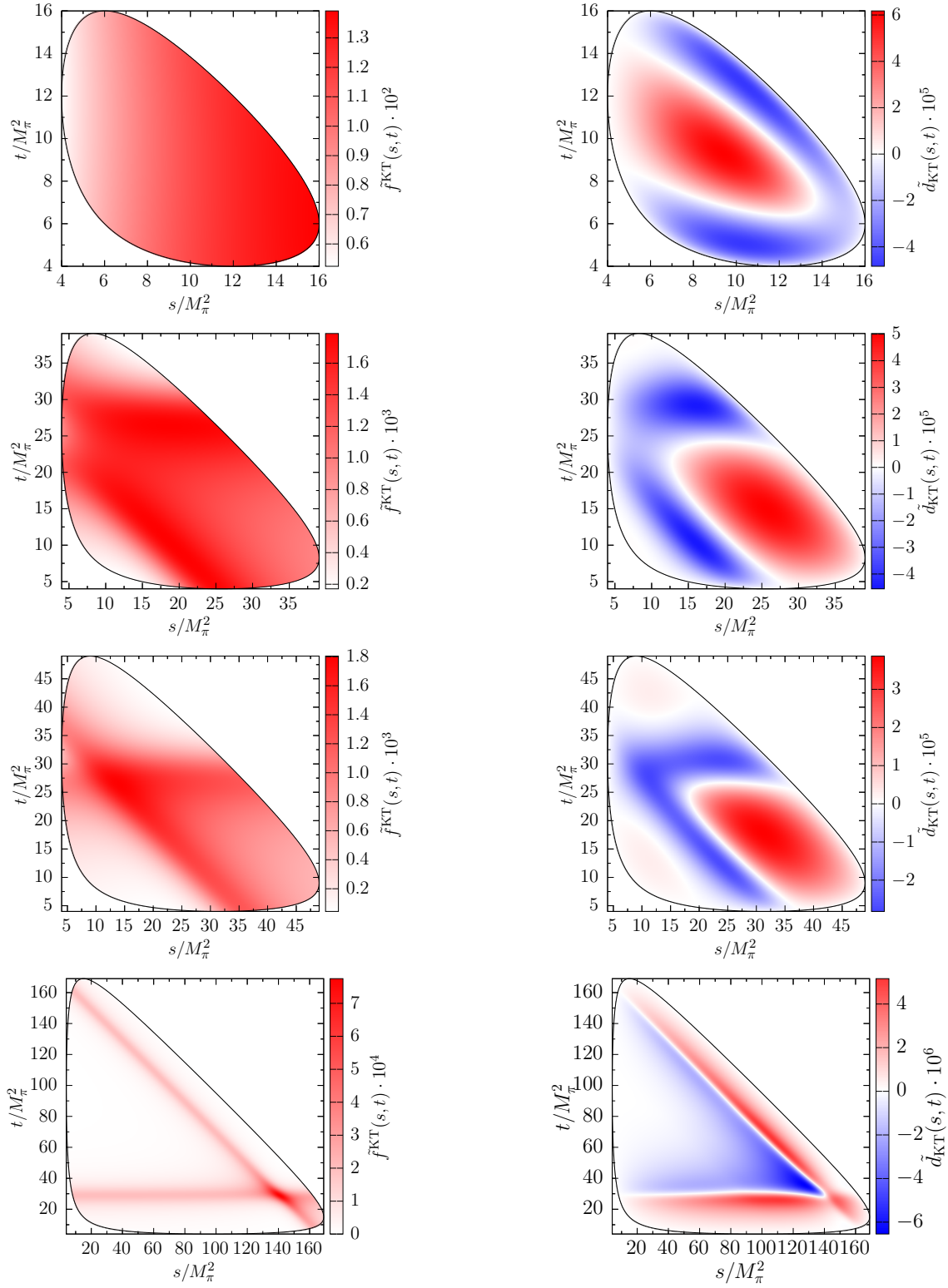

	\begin{subfigure}{0.49\textwidth}
	    \fontsize{12.5pt}{14pt} \selectfont
        \scalebox{0.67}{\input{plots/2++/dal_5.0.tex}}
	\end{subfigure}
	\begin{subfigure}{0.49\textwidth}
    	\fontsize{12.5pt}{14pt} \selectfont
        \scalebox{0.67}{\input{plots/2++/dal_comp_5.0.tex}}
	\end{subfigure}
	
	\begin{subfigure}{0.49\textwidth}
    	\fontsize{12.5pt}{14pt} \selectfont
        \scalebox{0.67}{\input{plots/2++/dal_7.25.tex}}
	\end{subfigure}
	\begin{subfigure}{0.49\textwidth}
        \fontsize{12.5pt}{14pt} \selectfont
        \scalebox{0.67}{\input{plots/2++/dal_comp_7.25.tex}}
	\end{subfigure}

	\begin{subfigure}{0.49\textwidth}
		\fontsize{12.5pt}{14pt} \selectfont
        \scalebox{0.67}{\input{plots/2++/dal_8.0.tex}}
	\end{subfigure}
	\begin{subfigure}{0.49\textwidth}
		\fontsize{12.5pt}{14pt} \selectfont
        \scalebox{0.67}{\input{plots/2++/dal_comp_8.0.tex}}
	\end{subfigure}
	
	\begin{subfigure}{0.49\textwidth}
		\fontsize{12.5pt}{14pt} \selectfont
        \scalebox{0.67}{\input{plots/2++/dal_14.0.tex}}
	\end{subfigure}
	\begin{subfigure}{0.49\textwidth}
		\fontsize{14pt}{14pt} \selectfont
        \scalebox{0.67}{\input{plots/2++/dal_comp_14.0.tex}}
	\end{subfigure}
	\caption{Dalitz plots for the $J^{PC}=2^{++}$ reconstruction theorem. From top to bottom the decay mass grows according to $M=5/7.25/8/14\,M_\pi$, while the Dalitz plot without the $\mathcal{K}(s,t,u)$ factor, $\Tilde{f}(s,t)$, defined via Eq.~\eqref{eq:pdf} is shown in the left and $\tilde{d}_\text{KL}(s,t)$ (Eq.~\eqref{eq:entropy_variance}) in the right column.} 
	\label{fig:dalitz_2pp}
\end{figure*}

For $0^{--}$ decays, we find some small numerical fluctuations in the low-mass region, which are due to the multiple regions where the total decay amplitude is kinematically suppressed in the Dalitz plots, shown in Fig.~\ref{fig:dalitz_0mm}: it vanishes along the three lines of $s=t$, $s=u$, and $t=u$. The number of events slightly rises up to approximately $N=10^6$ at the mass of the $\omega$ and stays constant up to $9M_\pi$. At higher masses, a slow decline sets in, including a small peak around $14M_\pi$. Due to the structure of the reconstruction theorem, the differences are located near the Dalitz plot boundaries. The six regions of intensity are then split into regions with larger pdfs for the KT equations and the Omn{\`e}s solutions, respectively. 

For $2^{++}$ decays, the low-mass behavior looks similar to the $1^{--}$ decays. The difference, however, is that we find a small plateau up to the $\omega$ mass, such that the number of events is increased by orders of magnitude. Furthermore the decline stops at the $\phi$ mass above $10^5$ events, such that the behavior at larger masses is more similar to the other quantum numbers. The Dalitz plots in Fig.~\ref{fig:dalitz_2pp} suggest that we do not have any visible features of the $\rho$ bands for the small decay masses. For larger masses we observe the two expected bands in $t$- and $u$-channel with one crossing point inside the Dalitz plot. Similar to the $1^{--}$ decays the Dalitz plot is split into two regions, where the pdfs for the KT equations and the Omn{\`e}s functions are of different size. These regions vary in shape for the different decay masses.

Finally, to test the dependence of our findings on the $\rho$ resonance width, we repeat the above exercises using a phase shift extracted from a simple Breit--Wigner model with an energy-dependent width~\cite{Ropertz:2018stk}, whose nominal width we fix to the smaller value $\Gamma_\rho=30\MeV$. We find that the number of events required to distinguish rescattering effects is significantly increased by about two orders of magnitude. This confirms the expected trend that rescattering effects vanish in the limit of small widths. A short analytic derivation of this limit is presented in \ref{app:narrow_res}.

\section{Conclusion} \label{sec:conclusions}

In this article, we have investigated the feasibility of unambiguously identifying the crossed-channel rescattering effects beyond the simplest isobar model in three-pion decays dominated by $\rho\pi$ intermediate states.  For four different quantum numbers of the decaying particles we have solved the Khuri--Treiman equations, integral equations that sum iterated two-pion rescattering in each pion pair to all orders.  We have determined the minimal sample sizes for which Dalitz-plot distributions allow us to distinguish the KT solutions from the naive picture that ignores all effects beyond two-body resonances.  The dependence on the mass of the decaying resonance has been studied in detail throughout.

The significance of the rescattering effects in $3\pi$ final states is heavily dependant on the decay kinematics. In particular, the appearance and position of the $\rho$ bands in the Dalitz plots plays a major role. For $J^{PC}=1^{--}$ we found a strong dependence on the mass of the decaying particle: rescattering effects are small for the $\omega$ resonance, where at least a few times $10^5$ events in a Dalitz plot are necessary to identify them at $5\sigma$ significance, while they are easily observable for the $\phi$ resonance, with of the order of $10^4$ Dalitz plot events sufficient; a similar sensitivity is expected for decaying isoscalar vector resonances up to almost $2\GeV$. However, the predictive power for large masses, e.g., the three-pion decays of the vector charmonia $J/\psi$ or $\psi'$~\cite{BESIII:2012vmy}, is clearly limited due to inelastic effects and higher partial waves.

In general, the processes with zeros in the amplitude due to crossing symmetry need more statistics to resolve rescattering effects.  This is the case for the isovector $3\pi$ system with quantum numbers $1^{-+}$ and $2^{++}$,\footnote{The $2^{++}$ system only has a zero in one of the two scalar functions.} as well as isoscalar $0^{--}$ states, where for a wide mass range up to $2\GeV$, we predict necessary statistics between $10^5$ and $10^6$ events to identify non-trivial rescattering effects at $5\sigma$ significance.  

Throughout, our investigation should be understood as a pilot study towards the thorough implementation of Dalitz plot fits beyond the simplest isobar models.  Theoretical limitations at this point clearly concern the constraint to Khuri--Treiman systems with one single free parameter: it is by no means guaranteed that the neglect of additional subtraction constants, which inter alia allow us to absorb effects of inelastic intermediate states, is justified in all circumstances.  Furthermore, decays with several relevant partial waves, in particular isoscalar $\pi\pi$ $S$-waves with their strong coupling to $K\bar K$ above $1\GeV$, pose additional difficulties to pin down corrections to two-pion lineshapes unambiguously; these are known to play an important role in the interpretation of resonance signals in the $a_1$ spectrum~\cite{COMPASS:2015kdx,Mikhasenko:2015oxp,COMPASS:2020yhb}.  The systematic study of such more complicated KT systems remains a both formidable and highly rewarding challenge for future research.

\begin{code}
\bsp
Generalized code for the solution of the Khuri--Treiman equations can be found at \href{https://github.com/HISKP-ph/khuri_treiman_solver}{github.com/HISKP-ph/khuri\_treiman\_solver}.
Some of the analysis of this article is documented at the webpage 
\href{https://mmikhasenko.github.io/rescattering-3pi-2023-001}{mmikhasenko.github.io/rescattering-3pi-2023-001}.
The exact versions are preserved at Refs.~\cite{KT_HISKP:2023,KTMC:2023}.
\esp
\end{code}

\begin{acknowledgements}
\bsp
We thank Fabian Krinner for collaboration in an early stage of this project.
Financial support by the DFG through the funds provided to the Sino--German Collaborative Research Center TRR110 ``Symmetries and the Emergence of Structure in QCD'' (DFG Project-ID 196253076 -- TRR 110) is gratefully acknowledged.
MM is funded by the Deutsche Forschungsgemeinschaft under Germany's Excellence Strategy -- EXC-2094 -- 390783311.
\esp
\end{acknowledgements}


\numberwithin{equation}{section}
\begin{appendix}

\section[1-+ reconstruction theorem]{$1^{-+}$ reconstruction theorem}\label{app:1mp-RT}

This appendix is dedicated to the derivation of the $1^{-+}$ reconstruction theorem using fixed-$t$ dispersion relations.
We start by considering the decay process
\begin{equation}
X^i(p)\to\pi^j(p_1)\pi^k(p_2)\pi^l(p_3)\,,
\end{equation}
where the $T$-matrix element is given by
\begin{align}
&\bra{\pi^j(p_1)\pi^k(p_2)\pi^l(p_3)}T\ket{X^i(p)}\notag\\
&\quad=(2\pi)^4\delta^{(4)}(p-p_1-p_2-p_3)\M^{ijkl}(s,t,u)\,,
\end{align}
and the Mandelstam variables are defined as
\begin{align}
s&=(p-p_1)^2=(p_2+p_3)^2\,,\notag\\
t&=(p-p_2)^2=(p_1+p_3)^2\,,\notag\\
u&=(p-p_3)^2=(p_1+p_2)^2\,.
\label{eq:mandelstam}
\end{align}
The decomposition into the scalar amplitude and the kinematic prefactor due to the odd intrinsic parity can be found in Eq.~\eqref{eq:amp_decomp}.
The isospin structure is identical to $\pi\pi$ scattering and the scalar amplitude therefore obeys the same decomposition
\begin{align}
\Ha^{ijkl}(s,t,u)&=\delta^{ij}\delta^{kl}\Ha_s(s,t,u)+\delta^{ik}\delta^{jl}\Ha_t(s,t,u)\notag\\
&\quad+\delta^{il}\delta^{jk}\Ha_u(s,t,u)\,.
\end{align}
Due to the symmetry of the process, the amplitude needs to stay invariant under simultaneous exchanges of isospin indices and momenta
\begin{align}
k&\leftrightarrow l \,, & p_2 &\leftrightarrow p_3 \,, & t &\leftrightarrow u\,;\notag\\
j&\leftrightarrow l\,, & p_1 &\leftrightarrow p_3\,, & s &\leftrightarrow u\,;\notag\\
j&\leftrightarrow k\, , & p_1 &\leftrightarrow p_2\,, & s &\leftrightarrow t\,,
\label{eq:transformation}
\end{align}
which relates $\Ha_s$, $\Ha_t$, and $\Ha_u$ and leads to
\begin{align}
\Ha^{ijkl}(s,t,u)&=\delta^{ij}\delta^{kl}\Ha(s,t,u)+\delta^{ik}\delta^{jl}\Ha(t,u,s)\notag\\
&\quad+\delta^{il}\delta^{jk}\Ha(u,s,t)\,,
\end{align}
where $\Ha$ is antisymmetric in its last two arguments.
The isospin projection operators are defined as
\begin{align}
\P_0^{ijkl}&=\frac{1}{3}\delta^{ij}\delta^{kl}\,,\notag\\
\P_1^{ijkl}&=\frac{1}{2}\left(\delta^{ik}\delta^{jl}-\delta^{il}\delta^{jk}\right)\,,\notag\\
\P_2^{ijkl}&=\frac{1}{2}\left(\delta^{ik}\delta^{jk}+\delta^{il}\delta^{jk}\right)-\frac{1}{3}\delta^{ij}\delta^{kl}\,,
\label{eq:isospin_proj}
\end{align}
which allow us to rewrite the isospin decomposition of the scalar amplitude according to
\begin{align}
\Ha^{ijkl}(s,t,u)&=\P_0^{ijkl}\Ha^0(s,t,u)+\P_1^{ijkl}\Ha^1(s,t,u)\notag\\
&\quad+\P_2^{ijkl}\Ha^2(s,t,u)\,,
\end{align}
resulting in
\begin{align}
\Ha(s,t,u)&=\frac{1}{3}\left(\Ha^0(s,t,u)-\Ha^2(s,t,u)\right)\,,\notag\\
\Ha(t,u,s)&=\frac{1}{2}\left(\Ha^1(s,t,u)+\Ha^2(s,t,u)\right)\,,\notag\\
\Ha(u,s,t)&=\frac{1}{2}\left(\Ha^2(s,t,u)-\Ha^1(s,t,u)\right)\,.
\end{align}
The partial-wave expansion of the isospin amplitudes proceeds via~\cite{Jacob:1959at}
\begin{equation}
\Ha^I(s,z_s)=\sum_{\ell=1}P_\ell'(z_s)a_\ell^I(s)\,.
\end{equation}
A fixed-$t$ dispersion relation for $\Ha(s,t,u)$ yields
\begin{align}
\Ha(s,t,u) &= P^t_{n-1}(s,u) \notag\\
&\quad + \frac{s^n}{2\pi i}\int_{4M_\pi^2}^\infty \dd s' \frac{\text{disc}_{s'} \Ha(s',t,u(s'))}{s'^n(s'-s)}\notag\\
&\quad + \frac{u^n}{2\pi i}\int_{4M_\pi^2}^\infty \dd u' \frac{\text{disc}_{u'} \Ha(s(u'),t,u')}{u'^n(u'-u)}\,,
\end{align}
where $n$ determines the number of subtractions and $P^t_{n-1}(s,u)$ is a polynomial in $s$ and $u$ of order $n-1$ with fixed $t$. We can insert the discontinuities according to
\begin{align}
\text{disc}_{s'}\,\Ha(s',t,u(s'))&=0\,,\notag\\
\text{disc}_{u'}\,\Ha(s(u'),t,u')&=\frac{1}{2}\disc a_1^1(u')\,,
\end{align}
neglecting all discontinuities in partial waves with $\ell \geq 2$.
If we employ fixed-$s$ and -$u$ dispersion relations in strict analogy to the above, we find that each fixed-variable dispersion relation misses the integral along the cut of the fixed Mandelstam variable. This missing integral can be subtracted from the polynomial if the number of subtraction constants is sufficiently high. This procedure is commonly referred to as symmetrizing the fixed-variable dispersion relations; details on general properties can be found in Ref.~\cite{Niehus:2022aau}. The result is given as
\begin{align}
\Ha(s,t,u)&=P_{n-1}(s,t,u)
-\frac{t^n}{4\pi i}\int_{4M_\pi^2}^\infty \dd t' \frac{\disc a_1^1(t')}{t'^n(t'-t)}\notag\\
&\qquad+\frac{u^n}{4\pi i}\int_{4M_\pi^2}^\infty \dd u' \frac{\disc a_1^1(u')}{u'^n(u'-u)}\,,
\end{align}
which can be simplified to
\begin{equation}
\Ha(s,t,u)=\Ha(t)-\Ha(u)\,,
\end{equation}
where
\begin{equation}
\Ha(s)=P_{n-1}(s)-\frac{s^n}{4\pi i}\int_{4M_\pi^2}^\infty \dd s' \frac{\disc a_1^1(s')}{s'^n(s'-s)}\,.
\end{equation}
The ambiguity of this decomposition is discussed in Sec.~\ref{sec:rt_1mp}.
Note that in order to consistently define the SVAs from the symmetrized version of $\Ha(s,t,u)$, $n\leq2$.
The same holds true in the cases of $1^{--}$ and $2^{++}$ (neglecting angular momenta beyond $P$-waves). 
For the $0^{--}$ decays, 
$n\leq 3$ subtractions can be implemented.

\section[2++ reconstruction theorem]{$2^{++}$ reconstruction theorem}\label{app:2pp-RT}
In this appendix we derive the reconstruction theorem for $J^{PC}=2^{++}$, applying the approximations needed for our analysis.
We start with the decay process
\begin{equation}
T^i(p)\to \pi^j(p_1)\pi^k(p_2)\pi^l(p_3)\,,
\end{equation}
where the Mandelstam variables are defined in Eq.~\eqref{eq:mandelstam}.
We can write down the amplitude in terms of two scalar amplitudes $\B$ and $\C$~\cite{Albaladejo:2019huw},
\begin{align}
\M^{ijkl}(s,t,u)&=\delta^{ij}\delta^{kl}\M_s(s,t,u)+\delta^{ik}\delta^{jl}\M_t(s,t,u)\notag\\
&\qquad+\delta^{il}\delta^{jk}\M_u(s,t,u)\notag\,,\\
\M_s(s,t,u)&=i\sqrt{2}\epsilon_{\mu\nu}K^\mu \left[(p_2+p_3)^\nu \B_s(s,t,u) \right.\notag\\
&\left.\qquad+(p_2-p_3)^\nu \C_s(s,t,u)\right]\notag\,,\\
\M_t(s,t,u)&=i\sqrt{2}\epsilon_{\mu\nu}K^\mu \left[(p_1+p_3)^\nu \B_t(s,t,u) \right.\notag\\
&\left.\qquad+(p_1-p_3)^\nu \C_t(s,t,u)\right]\notag\,,\\
\M_u(s,t,u)&=i\sqrt{2}\epsilon_{\mu\nu}K^\mu \left[(p_2+p_1)^\nu \B_u(s,t,u) \right.\notag\\
&\left.\qquad+(p_2-p_1)^\nu \C_u(s,t,u)\right]\,,
\label{eq:amplitude_a2}
\end{align}
with the isospin indices $i$, $j$, $k$, and $l$.
The amplitude needs to be invariant under the symmetry transformations in Eq.~\eqref{eq:transformation},
which gives us the following relations between the different $\B$ and $\C$ functions:
\begin{align}
\B_t(s,t,u)&=-\B_s(t,s,u)\,, & 
\C_t(s,t,u)&=-\C_s(t,s,u)\,,\notag\\
\B_u(s,t,u)&=-\B_s(u,t,s)\,, & 
\C_u(s,t,u)&=-\C_s(u,t,s)\,.
\end{align}
We find that $\B_s$ is antisymmetric in the last two arguments, while $\C_s$ is symmetric. This helps us to rewrite the amplitude in terms of $\B_s$ and $\C_s$, and we drop the subscript $s$ in the following.
The isospin projection operators are defined in Eq.~\eqref{eq:isospin_proj}.
The corresponding isospin amplitude can be decomposed in analogy to Eq.~\eqref{eq:amplitude_a2} using
\begin{align}
\M^{ijkl}(s,t,u)&=\P_0^{ijkl}\M^0(s,t,u)+\P_1^{ijkl}\M^1(s,t,u)\notag\\
&\qquad+\P_2^{ijkl}\M^2(s,t,u)\notag\,,\\
\M^I(s,t,u)&=i\sqrt{2}\epsilon_{\mu\nu}K^\mu \left[(p_2+p_3)^\nu \B^I(s,t,u) \right.\notag\\
&\left.\qquad+(p_2-p_3)^\nu \C^I(s,t,u)\right]\,,
\end{align}
where $I$ denotes the $s$-channel isospin.
This leads to the following relations including the scalar isospin amplitudes:
{\allowdisplaybreaks
\begin{align}
\B(s,t,u)&=\frac{1}{3}\left[\B^0(s,t,u)-\B^2(s,t,u)\right]\,,\notag\\
\C(s,t,u)&=\frac{1}{3}\left[\C^0(s,t,u)-\C^2(s,t,u)\right]\,,\notag\\
\B(t,s,u)&=\frac{1}{4}\left[\B^1(s,t,u)+\B^2(s,t,u)\right.\notag\\
&\qquad\left.+3\C^1(s,t,u)+3\C^2(s,t,u)\right]\,,\notag\\
\C(t,s,u)&=\frac{1}{4}\left[\B^1(s,t,u)+\B^2(s,t,u)\right.\notag\\
&\qquad\left.-\C^1(s,t,u)-\C^2(s,t,u)\right]\,,\notag\\
\B(u,t,s)&=\frac{1}{4}\left[-\B^1(s,t,u)+\B^2(s,t,u)\right.\notag\\
&\qquad\left.+3\C^1(s,t,u)-3\C^2(s,t,u)\right]\,,\notag\\
\C(u,t,s)&=\frac{1}{4}\left[\B^1(s,t,u)-\B^2(s,t,u)\right.\notag\\
&\qquad\left.+\C^1(s,t,u)-\C^2(s,t,u)\right]\,.
\end{align}}%
To find the relation between the scalar functions $\B(s,t,u)$, $\C(s,t,u)$ and partial-wave amplitudes,
we evaluate the isospin projected amplitude in the rest frame of the $s$-channel, resulting in
\begin{align}
\M^I_{\lambda = 0}(s,t,u) &= 0\,,\notag\\
\M^I_{\lambda = 1}(s,t,u) &= \beta(s) \bigg[ 
\B^I(s,t,u)\sin\theta_s \notag\\
&\quad -
\xi(s)\C^I(s,t,u) \sin\theta_s \cos\theta_s\bigg]\,,\notag\\
\M^I_{\lambda = 2}(s,t,u) &= \alpha(s) \C^I(s,t,u) \sin^2\theta_s\,,
\label{eq:M.explicit}
\end{align}
where the kinematic functions are defined as
\begin{align}
\alpha(s)&=-\frac{\lambda^{1/2}_T(s)\lambda_\pi(s)}{4\sqrt{2}\,s}\,, \qquad 
\beta(s)
=-\frac{\lambda_T(s)\lambda^{1/2}_\pi(s)}{8\sqrt{2s}\,M}\,,\notag\\
\xi(s)&=\sqrt{1-\frac{4M_\pi^2}{s}}\frac{s+M^2-M_\pi^2}{\lambda_T^{1/2}(s)}\,,
\end{align}
and
\begin{align}
\lambda_T(s)&=\lambda(s,M^2,M_\pi^2)\,,\quad
\lambda_\pi(s)=\lambda(s,M_\pi^2,M_\pi^2)\,.
\end{align}
Parity enforces the helicity-0 amplitude to vanish. We are therefore left with helicity-1 and helicity-2 amplitudes, whose partial-wave expansions start at $P$- and $D$-waves, respectively.
The $\M^I_\lambda$ contain kinematic singularities, but have well-defined partial-wave expansions for fixed isospin $I$:
\begin{align}
\M^I_\lambda(s,t,u)&=\sum_{j\geq |\lambda|}(2j+1)a^{\lambda I}_j(s)d_{\lambda 0}^j(z_s) \notag\\
&=\sum_{j\geq |\lambda|}(2j+1)K_{j\lambda}(s,t,u)\hat{a}^{\lambda I}_j(s)\hat{d}_{\lambda 0}^j(z_s) \,,\label{eq:PWE_general}
\end{align}
where in particular \cite{collins1977,Albaladejo:2019huw}
\begin{align}
K_{11}(s,t,u)&=\frac{1}{4\sqrt{s}}\sin\theta_s \lambda^{1/2}_\pi(s)\,,\notag\\
K_{21}(s,t,u)&=\frac{1}{4\sqrt{s}}\sin\theta_s \lambda_\pi(s)\lambda_T^{1/2}(s)\,,\notag\\
K_{22}(s,t,u)&=\frac{1}{16s}\sin^2\theta_s\lambda_\pi(s)\lambda_T^{1/2}(s)\,,
\end{align}
and
the Wigner $d$-matrices are given by
\begin{align}
d_{\lambda 0}^j(z_s) &= \hat{d}_{\lambda 0}^j(z_s)\sin^{|\lambda|}\theta_s\,, \quad
d_{10}^1(z_s)=-\frac{\sin\theta_s}{\sqrt{2}}\,,\notag\\
d_{10}^2(z_s)&=-\frac{3}{\sqrt{6}}\sin\theta_s\cos\theta_s \,,\quad d_{20}^2(z_s)=\frac{3}{2\sqrt{6}}\sin^2\theta_s\,.
\end{align}
Here, $\hat{a}^{\lambda I}_j(s)$ are the partial-wave amplitudes, which are however not yet free of kinematic constraints; see below.
We use the relation between $\M_1$, $\C$, and $\B$ to reduce $\B$ to its leading partial waves. Neglecting all higher ones (denoted by ellipses), we find
\begin{align}
\B^I&(s,t,u)\notag\\
&=\frac{1}{\beta(s)\sin\theta_s}  \M_1^I(s,t,u) + \xi(s)\cos\theta_s \C^I(s,t,u) \notag\\
&= \frac{6M}{\lambda_T(s)}\hat{a}^{1I}_1(s) + \frac{16}{3\sqrt{3}}\lambda_\pi^{1/2}(s)\lambda_T^{1/2}(s)z_s\hat{a}^{1I}_2(s)\notag\\
&\quad+\xi(s)z_s \C^I(s,t,u) +
\ldots \notag\\
&\equiv \Tilde{a}^{1I}_1(s) + \chi(s) z_s \Tilde{a}^{1I}_2(s) + \xi(s) z_s \C^I(s,t,u) + \ldots\,. \label{eq:PWE}
\end{align}
Note that a kinematic constraint needs to be enforced on $\hat{a}^{1I}_1$ to cancel the zeros of $\lambda_T(s)$.  As a consequence, $\tilde{a}^{1I}_1$ is now a partial wave free of any kinematic singularities and zeros, and therefore apt for a generalized Omn{\`e}s representation.
In strict analogy $\C^I$ can be expressed via 
\begin{align}
\C^I(s,t,u)&=\frac{1}{\alpha(s)\sin^2\theta_s}\M^I_2(s,t,u)=-\frac{5\sqrt{3}}{8}\hat{a}_2^{2I}(s)\notag\\
&\equiv -\Tilde{a}_2^{2I}(s)\,,
\end{align}
employing the partial-wave expansion of $\M_2^I$.  
Using that the $P$-waves are pure isospin $I=1$ and $D$-waves need to be $I=0$ or $I=2$, we can give the relations for the discontinuities of $\B$ and $\C$, e.g.,
\begin{align}
\text{disc}_{s'}\,& \B(s',t,u(s')) \notag\\
&= \frac{1}{3} \big(\chi(s')z_{s'} \disc \Tilde{a}^{10}_2(s') - \xi(s')z_{s'}\disc \Tilde{a}^{20}_2(s')\notag\\
&\quad -\chi(s')z_{s'} \disc \Tilde{a}^{12}_2(s') + \xi(s')z_{s'}\disc \Tilde{a}^{22}_2(s')\big)\,,\notag\\
\text{disc}_{u'}\,& \B(s(u'),t,u') \notag\\
&= \frac{1}{4}\big(-\disc \Tilde{a}^{11}_1(u')+3\disc \Tilde{a}^{22}_2(u')\notag\\
&\quad +\chi(u')z_{u'} \disc \Tilde{a}^{12}_2(u') - \xi(u')z_{u'} \disc \Tilde{a}^{22}_2(u')\big)\,.
\end{align}
Writing down a fixed-$t$ dispersion relation for $\B$
\begin{align}
\B(s,t,u) &= P^t_{n-1}(s,u) \notag\\
&\quad + \frac{s^n}{2\pi i}\int_{4M_\pi^2}^\infty \dd s' \frac{\text{disc}_{s'}\, \B(s',t,u(s'))}{s'^n(s'-s)}\notag\\
&\quad + \frac{u^n}{2\pi i}\int_{4M_\pi^2}^\infty \dd u' \frac{\text{disc}_{u'}\, \B(s(u'),t,u')}{u'^n(u'-u)}\,,
\end{align}
and inserting the expansion from Eq.~\eqref{eq:PWE} into partial waves in fixed-$s$, -$t$, and -$u$ dispersion relations and symmetrizing these equations yields
\begin{align}
\B(s,t,u)&=\B_1(t)-\B_1(u)+(u-s)\B_2(t)\notag\\
&\qquad+(s-t)\B_2(u)-3\C_2(t)+3\C_2(u)\notag\\
&\qquad+(t-u)\left(\B_0(s)-\frac{4}{3}\B_2(s)\right)\,,\notag\\
\C(s,t,u)&=\B_1(t)+\B_1(u)+(u-s)\B_2(t)\notag\\
&\qquad-(s-t)\B_2(u)+\C_2(t)+\C_2(u)\notag\\
&\qquad-\left(\C_0(s)-\frac{4}{3}\C_2(s)\right)\,,
\end{align}
where
\begin{align}
\B_1(s)&=P^{\B1}_{n-1}(s)+\frac{s^n}{8 \pi i} \int_{4M_\pi^2}^\infty \dd s' \frac{\disc \Tilde{a}_1^{11}(s')}{s'^n(s'-s)}\,,\notag\\
\B_0(s)&=P^{\B0}_{n-2}(s)\notag\\
+&\frac{s^{n-1}}{6 \pi i} \int_{4M_\pi^2}^\infty \dd s' \frac{\chi(s')\disc \Tilde{a}_2^{10}(s')-\xi(s')\disc \Tilde{a}_2^{20}(s')}{\kappa(s')s'^{n-1}(s'-s)}\,,\notag\\
\B_2(s)&=P^{\B2}_{n-2}(s)\notag\\
+&\frac{s^{n-1}}{8 \pi i} \int_{4M_\pi^2}^\infty \dd s' \frac{\chi(s')\disc \Tilde{a}_2^{12}(s')-\xi(s')\disc \Tilde{a}_2^{22}(s')}{\kappa(s')s'^{n-1}(s'-s)}\,,\notag\\
\C_0(s)&=P^{\C0}_{n-1}(s)+\frac{s^n}{6 \pi i} \int_{4M_\pi^2}^\infty \dd s' \frac{\disc \Tilde{a}_2^{20}(s')}{s'^n(s'-s)}\,,\notag\\
\C_2(s)&=P^{\C2}_{n-1}(s)+\frac{s^n}{8 \pi i} \int_{4M_\pi^2}^\infty \dd s' \frac{\disc \Tilde{a}_2^{22}(s')}{s'^n(s'-s)}\,.
\end{align}
At this point we neglect all SVAs that contain discontinuities for $D$-waves. Therefore only $\B_1$ is left, which contains the $I=1$ $P$-wave. The inhomogeneity is defined via
\begin{equation}
\B_1(s)+\widehat{\B}_1(s)=\frac{1}{4}\Tilde{a}_1^{11}(s)\,.
\end{equation}
We can find the partial wave by projecting Eq.~\eqref{eq:PWE}:
\begin{align}
\Tilde{a}_1^{1I}(s)&=\frac{3}{4}\int_{-1}^1\dd z_s (1-z_s^2)\big(\B^I(s,t,u)\notag\\
&\qquad\qquad-\xi(s)z_s\C^I(s,t,u)\big)\,,
\end{align}
and therefore
\begin{equation}
\widehat{\B}_1(s)=\frac{3}{4}\left[\langle(1-z_s^2)\B_1\rangle - \xi(s) \langle(1-z_s^2)z_s\B_1\rangle\right]\,,
\end{equation}
by inserting
\begin{align}
\B^1(s,t,u)&=\frac{1}{2}\big[\B(t,s,u)+3\C(t,s,u)\notag\\
&\qquad\qquad-\B(u,t,s)+3\C(u,t,s)\big]\notag\\
&=4\B_1(s)+\B_1(t)+\B_1(u)\notag\,,\\
\C^1(s,t,u)&=\frac{1}{2}\big[\B(t,s,u)-\C(t,s,u)\notag\\
&\qquad\qquad+\B(u,t,s)+\C(u,t,s)\big]\notag\\
&=\B_1(t)-\B_1(u)\,.
\end{align}

\section{\texorpdfstring{$D$}{D}-wave projection}
This appendix is an extension of the numerical treatment of Khuri--Treiman amplitudes for $S$- and $P$-waves~\cite{Niecknig2011,Schneider2012,Isken2015} to $D$-waves. Even though we do not introduce $D$-wave discontinuites in our analysis, the singularity structure of the inhomogeneity integral in the $2^{++}$ reconstruction theorem is of this type. 
The projection integrals from \ref{app:2pp-RT} can be written in the form
\begin{equation}
\langle z_s^n \B \rangle = \frac{2^n}{\kappa^{n+1}(s)}\int_{s_{-}(s)}^{s_+(s)}\dd s' (s'-\sigma)^n \B(s')\,,
\end{equation}
where
\begin{align}
\sigma &= \frac{1}{2}(3s_0-s)\,, & 
z_s &= \frac{2}{\kappa(s)}(s'-\sigma)\,,\notag\\
s_\pm(s)&=\frac{1}{2}(3s_0-s\pm\kappa(s))\,.
\end{align}
Therefore the projection integral has a maximal singularity of degree $5/2$ at the pseudothreshold $s_\text{III}=(M-M_\pi)^2$.\footnote{Note the additional factor from the $\xi$ function.} The goal is to tame the effect of this singularity. Thereby we rewrite the integrals into pieces that are analytically solvable and ones which are numerically stable. We start by explicitly rewriting
\begin{equation}
\Tilde{\B}(s)=\kappa^5(s)\widehat{\B}(s)\,,
\end{equation}
which leads to the following form of the dispersive integral:
\begin{equation}
I(s)=\int_{\sth}^\infty \frac{\dd s'}{s'^n}\frac{\sin\delta(s')\Tilde{\B}(s')}{\kappa^5(s')|\Omega(s')|(s'-s\mp i \epsilon)}\,.
\end{equation}
The singularities at the scattering thresholds $\sth=4M_\pi^2$ and $\sIV=(M+M_\pi)^2$ are removable, since the integration path at these values is point-like. Furthermore we define a function $\nu(s)$ to account for the correct analytic continuation:
\begin{align}
\nu(s)=\begin{cases}\sqrt{1-\displaystyle\frac{\sth}{s}}\sqrt{\sIV-s}&\text{for } s<\sIV\\
i\sqrt{1-\displaystyle\frac{\sth}{s}}\sqrt{s-\sIV}&\text{for }s>\sIV\end{cases}\,.
\end{align}
We keep the singularity at the pseudothreshold explicit in the dispersion integral and define a function $T(s)$ containing all other functions,
\begin{equation}
T(s)=\frac{\sin\delta(s)\Tilde{\B}(s)}{s^n\nu^5(s)|\Omega(s)|}\,.
\end{equation}
The dispersion integral is then given by
\begin{equation}
I(s)=\int_{\sth}^\infty \dd s' \frac{T(s')}{(\sIII-s')^{5/2}(s'-s\mp i \epsilon)}\,.
\end{equation}
Using this form we present the procedure to handle the Cauchy and the pseudothreshold singularities for $D$-waves. This is achieved by splitting the integrals into analytic parts and numerical integrals with removable singularities. In the following we consider two different regions for $s$:
\begin{enumerate}
    \item $s\in \mathbb{R} \wedge s<\sth$ or $s\in\mathbb{C}$
    
    Here $s$ cannot hit the Cauchy singularity, since it is outside of the integration range. Therefore we only need to handle the critical point $s'=\sIII$.  For this purpose we add and subtract an expansion of $T$ up to the second derivative around the pseudothreshold:
    \begin{align}
    I(s)&=\int_{\sth}^{\Lambda^2} \dd s' \frac{\Tilde{T}(s')}{(\sIII-s')^{5/2}(s'-s)}\notag\\
    &\quad+T(\sIII)\Q_{5/2}(s,\sth,\Lambda^2)\notag\\
    &\quad+T'(\sIII)\Q_{3/2}(s,\sth,\Lambda^2)\notag\\
    &\quad+T''(\sIII)\Q_{1/2}(s,\sth,\Lambda^2)\,,\notag\\
    \Tilde{T}(s)&=T(s)-T(\sIII)-(\sIII-s)T'(\sIII)\notag\\
    &\quad-(\sIII-s)^2T''(\sIII)\,.
    \end{align}
    The procedure to calculate the derivatives is explained at the end of this appendix. Furthermore we introduce a high-energy cutoff $\Lambda^2$.
    In this form the numerical integral converges and the $\Q$ functions are analytically given by
    \begin{align}
    \Q_{\frac{2n+1}{2}}(s,x,y)&=\int_x^y\frac{\dd s'}{\sqrt{\sIII-s'}^{2n+1}(s'-s)}\notag\\
    =\frac{1}{\sIII-s}&\Bigg(\frac{-2i}{(2n-1)\sqrt{y-\sIII}^{2n-1}}\notag\\
    &\quad-\frac{2}{(2n-1)\sqrt{\sIII-x}^{2n-1}}\notag\\
    &\quad+\Q_{\frac{2n-1}{2}}(s,x,y)\Bigg)\quad \forall n\in\mathbb{N}_{>0}\notag\,,
    \end{align}
    \begin{align}
    \Q_{\frac{1}{2}}(s,x,y)&=\frac{1}{\sqrt{\sIII-s}}\notag\\
    &\quad\cdot\Bigg(\log\frac{\sqrt{\sIII-s}+\sqrt{\sIII-x}}{\sqrt{\sIII-s}-\sqrt{\sIII-x}}\notag\\
    &\qquad-2i\arctan\frac{\sqrt{y-\sIII}}{\sqrt{\sIII-s}}\Bigg)\,.
    \end{align}
    This recursive relation is proven by induction.
    
    \item $s\in \mathbb{R} \wedge s>\sth$

    Here we introduce an artificial cutoff $p=(\sIII+s)/2$ to separate the two singularities into different integrals.
    First, we investigate $s<\sIII$, where the Cauchy singularity at $s'=s$ is in the integral from $\sth$ to $p$ and the pseudothreshold singularity is in the integral from $p$ to $\Lambda^2$. We therefore rewrite the integral in the following form:
    \begin{align}
    I(s)&=\int_{\sth}^p\dd s' \frac{T(s')-T(s)}{(\sIII-s')^{5/2}(s'-s)}\notag\\
    &\quad+T(s)\R_{5/2}(s,\sth,p)\notag\\
    &\quad+\int_p^{\Lambda^2}\dd s' \frac{\Tilde{T}(s')}{(\sIII-s')^{5/2}(s'-s)}\notag\\
    &\quad+T(\sIII)\Q_{5/2}(s,p,\Lambda^2)\notag\\
    &\quad+T'(\sIII)\Q_{3/2}(s,p,\Lambda^2)\notag\\
    &\quad+T''(\sIII)\Q_{1/2}(s,p,\Lambda^2)\,,
    \end{align}
    where the analytic form of $\R$ is given by
    \begin{align}
    \R_{\frac{2n+1}{2}}(s,x,y)&=\int_x^y\frac{\dd s'}{\sqrt{\sIII-s'}^{2n+1}(s'-s\mp i \epsilon)}\notag\\
    =\frac{1}{\sIII-s}&\Bigg(\frac{2}{(2n-1)\sqrt{\sIII-y}^{2n-1}}\notag\\
    &\quad-\frac{2}{(2n-1)\sqrt{\sIII-x}^{2n-1}}\notag\\
    &\quad+\R_{\frac{2n-1}{2}}(s,x,y)\Bigg)\quad \forall n\in\mathbb{N}_{>0}\notag\,,
    \end{align}
    \begin{align}
    \R_{\frac{1}{2}}(s,x,y)&=\frac{1}{\sqrt{\sIII-s}}\notag\\
    &\quad\cdot\Bigg(\log\frac{\sqrt{\sIII-x}+\sqrt{\sIII-s}}{\sqrt{\sIII-x}-\sqrt{\sIII-s}}\notag\\
    &\qquad+\log\frac{\sqrt{\sIII-s}-\sqrt{\sIII-y}}{\sqrt{\sIII-s}+\sqrt{\sIII-y}}\pm i\pi \Bigg)\,.
    \end{align}
    These equations are again proven by induction.

    Next we investigate the case $s>\sIII$, where the Cauchy singularity at $s'=s$ is in the integral from $p$ to $\Lambda^2$ and the pseudothreshold singularity is in the integral from $\sth$ to $p$. Therefore the integral can be rewritten as 
    \begin{align}
    I(s)&=\int_{p}^{\Lambda^2}\dd s' \frac{T(s')-T(s)}{(\sIII-s'\mp i\epsilon)^{5/2}(s'-s)}\notag\\
    &\quad+T(s)\R_{5/2}(s,p,\Lambda^2)\notag\\
    &\quad+\int_{\sth}^{p}\dd s' \frac{\Tilde{T}(s')}{(\sIII-s')^{5/2}(s'-s)}\notag\\
    &\quad+T(\sIII)\Q_{5/2}(s,\sth,p)\notag\\
    &\quad+T'(\sIII)\Q_{3/2}(s,\sth,p)\notag\\
    &\quad+T''(\sIII)\Q_{1/2}(s,\sth,p)\,.
    \end{align}
\end{enumerate}
The remaining integrals can be evaluated numerically. We therefore expand $T(s)$ in a series around the pseudothreshold:
\begin{align}
T(s)&=T(\sIII)+(\sIII-s)T'(\sIII)\notag\\
&\quad+(\sIII-s)^2T''(\sIII) +\sqrt{\sIII-s}^5 d_D\notag\\
&\quad+ (\sIII-s)^3 e_D + \ldots\notag\\
&=a_D + b_D (\sIII-s) + c_D (\sIII-s)^2 \notag\\
&\quad+d_D \sqrt{\sIII-s}^5  + e_D (\sIII-s)^3  + \ldots\notag\,,\\
\frac{\Tilde{T}(s)}{\sqrt{\sIII-s}^5} &= d_D  + e_D \sqrt{\sIII-s}  + \ldots \,.
\end{align}
This expansion is matched both slightly above and below the pseudothreshold. For each of these the five constraints are fixed by the function at pseudothreshold $\sIII$ and two matching points $\sIII\pm\epsilon$ and $\sIII\pm 4\epsilon$ as well as the first and second derivatives at the matching point $\sIII\pm\epsilon$. The parameters are then evaluated via
\begin{align}
f(x,i,j,k,l,m,n,p)&=-\frac{1}{n\epsilon^p}\big(iT(x)+jT(x-3\epsilon)\notag\\
&+kT(\sIII)+l\epsilon T'(x) 
+m\epsilon^2 T''(x)\big)
\end{align}
and
\begin{align}
a_D&=T(\sIII) \,,\notag\\
b_D&=f(x,-112,1,111,-80,-32,28,1)\big|_{x=\sIII-\epsilon} \,,\notag\\
c_D&=f(x,126,-3,-123,114,68,14,2)\big|_{x=\sIII-\epsilon} \,,\notag\\
d_D&=-2f(x,28,-1,-27,24,18,7,5/2)\big|_{x=\sIII-\epsilon} \,,\notag\\
e_D&=f(x,56,-3,-53,44,40,28,3)\big|_{x=\sIII-\epsilon}
\end{align}
below pseudothreshold and
\begin{align}
a_D&=T(\sIII) \,,\notag\\
b_D&=f(x,112,-1,-111,-80,32,28,1)\big|_{x=\sIII+\epsilon} \,,\notag\\
c_D&=f(x,126,-3,-123,-114,68,14,2)\big|_{x=\sIII+\epsilon} \,,\notag\\
d_D&=-2if(x,28,-1,-27,-24,18,7,5/2)\big|_{x=\sIII+\epsilon} \,,\notag\\
e_D&=f(x,-56,3,53,44,-40,28,3)\big|_{x=\sIII+\epsilon}
\end{align}
above the pseudothreshold.

\section{Interference ring}\label{app:interferencering}

\begin{figure*}
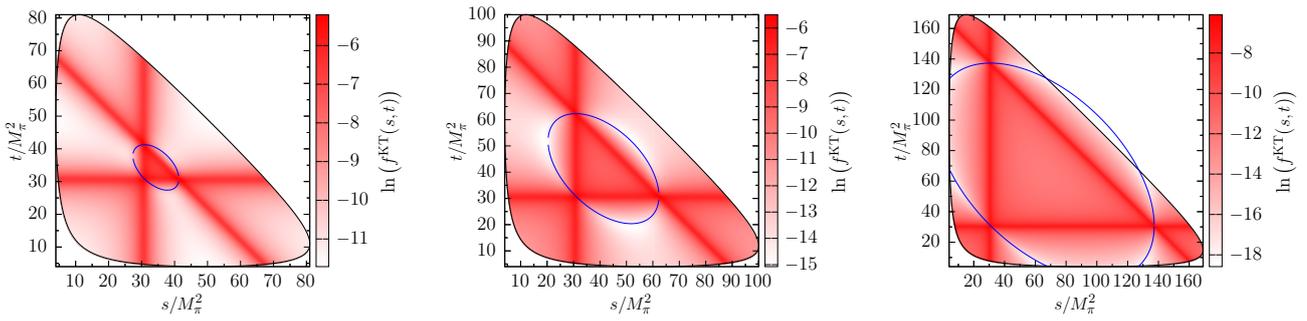

    \hspace{-0.8cm}
	\begin{subfigure}{0.33\textwidth}
	    \fontsize{13pt}{14pt} \selectfont
        \scalebox{0.55}{\input{plots/1--_small/dal_10.0.tex}}
	\end{subfigure}
	\begin{subfigure}{0.33\textwidth}
	    \fontsize{13pt}{14pt} \selectfont
        \scalebox{0.55}{\input{plots/1--_small/dal_11.0.tex}}
	\end{subfigure}
	\begin{subfigure}{0.33\textwidth}
	    \fontsize{13pt}{14pt} \selectfont
        \scalebox{0.55}{\input{plots/1--_small/dal_14.0.tex}}
	\end{subfigure}
	\caption{Logarithm of the intensity in the Dalitz plots for $J^{PC}=1^{--}$ with smaller $\rho$ width, see \ref{app:interferencering}. From left to right, we show the decay masses $M=10/11/14\,M_\pi$. The blue line is the solution of Eq.~\eqref{eq:inter_param}.}
	\label{fig:dalitz_1mm_inter}
\end{figure*}

When studying the Dalitz plots of the $1^{--}$ decays, we find a remarkable property induced by the symmetry of the process: a ring-shaped local minimum in the logarithmic intensity that crosses all three intersection points of the $\rho$ bands. However, this only becomes visible when using an unphysically narrow $\rho$ resonance. We again (\cf\  Sec.~\ref{sec:results}) employ a phase shift from a simple Breit--Wigner model with the energy-dependent width from Ref.~\cite{Ropertz:2018stk} and the nominal width $\Gamma_\rho=30\MeV$; therefore the resonance bands in the Dalitz plots are much narrower. When looking at plots with $M^2>3(M_\rho^2-M_\pi^2)$, we obtain an interference ring that crosses all three intersection points of the three $\rho$ bands. Transforming to the well-known Dalitz plot variables~\cite{Dalitz:1953cp,Fabri:1954zz,Weinberg:1960zza}
\begin{align}
x&=\frac{\sqrt{3}}{2MQ}(t-u)\,,\notag\\
y&=\frac{3}{2MQ}\left((M-M_\pi)^2-s\right)-1\,,\notag\\
Q&=M-3M_\pi
\end{align}
renders the ring to a perfect circle. Using the three intersection points $s=t=M_\rho^2$, $s=u=M_\rho^2$, and $t=u=M_\rho^2$, we obtain three equations of the form
\begin{equation}
(x-x_c)^2+(y-y_c)^2-r^2=0\,.
\end{equation}
These can be solved for the center and the radius of the circle in this parameterization:
\begin{align}
x_c&=0\,,\qquad y_c=0\,,\notag\\
r^2&=\left(\frac{3(M_\rho^2-M_\pi^2)-M^2}{MQ}\right)^2\,.
\end{align}
Reverting back to the Mandelstam variables, we find that the ring can be determined using the formula for the circle $x^2+y^2-r^2=0$ and solving for
\begin{align}
t(s)&=\frac{1}{2}\left(M^2+3M_\pi^2-s\pm\sqrt{M_1(s)M_2(s)}\right)\,,\notag\\
M_1(s)&=M^2+3M_\pi^2-s-2M_\rho^2\notag\,,\\
M_2(s)&=M^2+3(M_\pi^2+s-2M_\rho^2)\,,\label{eq:inter_param}
\end{align}
in the $s$ domain
\begin{align}
-\frac{M^2}{3}-M_\pi^2+2M_\rho^2\leq s\leq M^2+3M_\pi^2-2M_\rho^2\,.
\end{align}
This allows us to plot the interference ring in the Dalitz plot (\cf\  Fig.~\ref{fig:dalitz_1mm_inter}). For large masses, a sizeable part of it is outside of the Dalitz plot or close to its boundary, where the phase space is small. Using our realistic parameterization for the $\rho$ resonance, 
this feature is washed out by the broadness of the $\rho$ resonance.

\section{Rescattering effects for narrow resonances}\label{app:narrow_res}

We assume that for a narrow-width resonance, the Omn{\`e}s function behaves approximately like a Breit--Wigner parameterization~\cite{Breit:1936zzb}
\begin{equation}
\Omega(s)=\frac{M^2}{M^2-s-iM\Gamma}\,,\label{eq:BW}
\end{equation}
where the phase is given by
\begin{equation}
\delta(s)=\arctan\left(\frac{M\Gamma}{M^2-s}\right)\,.\label{eq:BW_phase}
\end{equation}
This assumption is justified since a zero-width phase given by
\begin{equation}
\delta(s)=\pi \theta(s-M^2)\,.
\end{equation}
leads to 
\begin{equation}
\Omega(s)=\frac{M^2}{M^2-s}\,,
\end{equation}
which is a Breit--Wigner function with zero width.
By using Eqs.~\eqref{eq:BW} and \eqref{eq:BW_phase}, the phase-shift-dependent fraction inside the dispersion integral over the inhomogeneity in Eq.~\eqref{eq:inhomOmnes-sol} reads
\begin{equation}
\frac{\sin\delta(s)}{|\Omega(s)|}=\frac{\Gamma}{M}\,.
\end{equation}
The general KT solution~\eqref{eq:inhomOmnes-sol} therefore reduces to
\begin{align}
\X(s)&=\Omega(s)\Bigg(P_{n-1}(s)+\frac{\Gamma}{M}\cdot\frac{s^n}{\pi}\int_{4M_\pi^2}^\infty \frac{\dd s'}{s'^n}\frac{\widehat{\X}(s')}{(s'-s)}\Bigg)\,,
\end{align}
making it amply clear that all crossed-channel rescattering effects are suppressed in the limit of a narrow-width resonance for $\Gamma\to 0$. 

\end{appendix}

\bibliographystyle{utphysmod}
\bibliography{Literature}

\end{document}